\documentclass[12pt]{article}

\usepackage{macros}
\usepackage{epsf,amsfonts,hyperref}
\renewcommand\L{\Lambda}
\newcommand\T{{\cal T}}

\newcommand\Res{\mathop{\rm Res}}
\newcommand\Pol{\mathop{\rm Pol}}
\newcommand\ubar{\overline{u}}
\newcommand\Zbar{\overline{Z}}
\newcommand\Fbar{\overline{F}}

\newcommand\ps{\psi}
\newcommand\hatps{\phi}

\begin{document}
\sloppy


\pagestyle{empty}
\hfill SPT-06/046
\addtolength{\baselineskip}{0.20\baselineskip}
\begin{center}
\vspace{26pt}
{\large \bf {Universal distribution of random matrix eigenvalues near the ``birth of a cut'' transition.}}
\newline
\vspace{26pt}

{\sl B.\ Eynard}\hspace*{0.05cm}\footnote{ E-mail: eynard@spht.saclay.cea.fr
}\\
\vspace{6pt}
Service de Physique Th\'{e}orique de Saclay,\\
F-91191 Gif-sur-Yvette Cedex, France.\\
\end{center}

\vspace{20pt}
\begin{center}
{\bf Abstract}
\end{center}
%

We study the eigenvalue distribution of a random matrix, at a transition where a new connected component of the eigenvalue density support
appears away from other connected components. Unlike previously studied critical points, which correspond to rational singularities $\rho(x)\sim x^{p/q}$ classified by conformal minimal models and integrable hierarchies, this transition shows logarithmic and non-analytical behaviours. There is no critical exponent, instead, the power of $N$ changes in a saw teeth behaviour.





\pagestyle{plain}
\setcounter{page}{1}


\section{Introduction}

Random matrix models \cite{Mehta, ZJDFG} have been studied in relationship with many areas of
physics and mathematics.
The reason of their success for most of their applications is their
``universality'' property, i.e. the fact that the eigenvalues statistical
distribution of a large random matrix depends only on the symmetries of
the matrix ensemble, and not on the detailed Boltzmann weight (characterized by
a potential).
Although this universality property has been much studied for generic
potentials, some universality should also hold for critical potentials.
Different kinds of critical potentials have been studied, and their
universality classes have been found to be in correspondence with
non-linear integrable hierarchies (KdV, MKdV, KP,...) \cite{BlIt,BlEycrit, CDM, PeS, Cicuta, DDJT},
and with the $(p,q)$ rational minimal models of conformal field theory \cite{KazakoVDK}.
They correspond to rational singularituies of the equilibrium density:
\beq
\rho(x) \sim (x-a)^{p/q}\, .
\eeq
In the hermitian 1-matrix model, we have only $q=2$ and $p$ arbitrary, thus we get only half integer singularities
(hyperelliptical curves), which are in relationship with the KdV hierarchies,
whereas a 2-matrix model allows to have any rational singularity $(p,q)$ \cite{KazakoVDK}.
The specific heat near such a rational singularity obeys a Gelfand-Dikii-type equation (Painlev\'e I equation for $(p,q)=(3,2)$).
A well known case is the edge of the spectrum where $(p,q)=(1,2)$, which gives Tracy-Widom law \cite{TW}, and which is governed by the Painlev\'e II equation.
Another well known case is the merging of two cuts (Bleher and Its \cite{BlIt,BI2}),
where $(p,q)=(2,1)$, which is also governed by a Painlev\'e II equation as well (indeed, $(p,q)$ and $(q,p)$ are known to be dual to each other \cite{DFMS}).

\medskip

Here, we shall study a kind of critical point which has been mostly disregarded
(because usual methods don't apply to it): ``the birth of a cut critical point''\footnote{Name suggested by P. Bleher who initiated this work.}.

\begin{figure}[bth]
\hrule\hbox{\vrule\kern8pt
\vbox{\kern8pt \vbox{
\begin{center}
{\mbox{\epsfxsize=10.truecm\epsfbox{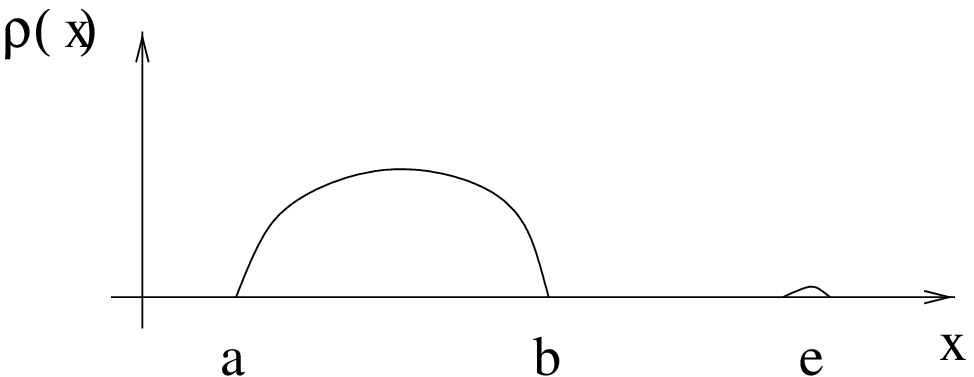}}}
\end{center}
\caption{The ``birth of a cut'' density of eigenvalues is such that
one of the connected components of the support contains a very small number
of eigenvalues ($\ll n$).\label{figrhocrit}}
}\kern8pt}
\kern8pt\vrule}\hrule
\end{figure}

Such a critical point, is characterized by the fact that when
a parameter of the model (let us call it temperature) is varied, a new connected component
appears in the support of the large $N$ average eigenvalue density.
When the temperature $T$ is just above critical temperature $T_c$,
the number of eigenvalues in the newborn connected component is small (see fig.\ref{figrhocrit}),
and thus, many usual large $n$ methods don't work in that case.

Our goal is to study the eigenvalues statictics in the vicinity of the
critical point, and find its universality class.

In this purpose, we shall start from the partition function,
and treat the eigenvalues in the other cuts with mean field approximations,
and reduce the problem to an effective partition function for
eigenvalues in the newborn cut only, in a method similar to \cite{BDE}.

We find that, unlike rational critical points, the birth of a cut critical point does not corrsepond to power law behaviors or transcandental differential equations, but it exhibits logarithmic behaviors, and discontinuous functions.

\bigskip

\begin{figure}[bth]
\hrule\hbox{\vrule\kern8pt
\vbox{\kern8pt \vbox{
\begin{center}
{\mbox{\epsfxsize=10.truecm\epsfbox{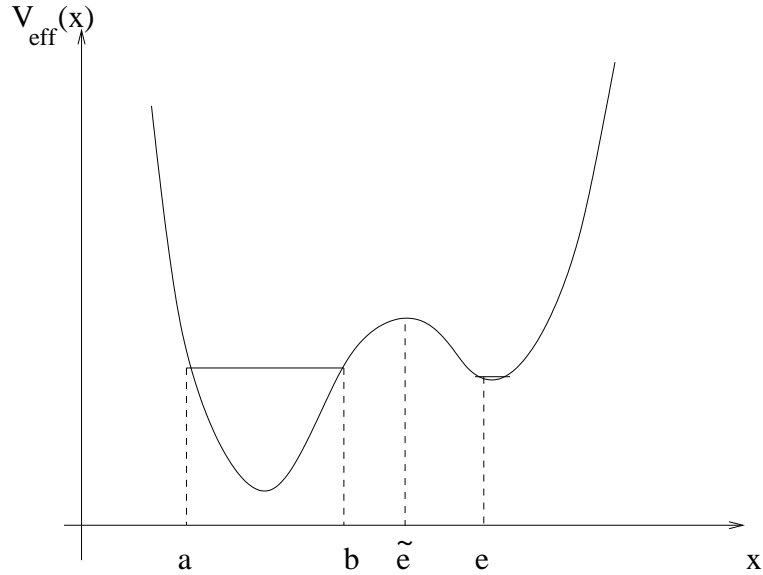}}}
\end{center}
\caption{The ``birth of a cut'' critical potential is such that one of the
potential wells of the effective potential is just at the Fermi level.
At a $\nu^{\rm th}$ order critical point, the effective potential behaves like
$V_{{\rm eff}}(e)+ {1\over 2\nu !}V^{(2\nu)}_{{\rm eff}}(e)\, (x-e)^{2\nu}+\dots$.\label{figpotcrit}}
}\kern8pt}
\kern8pt\vrule}\hrule
\end{figure}

The matrix model is associated to a family of orthogonal polynomials, whose zeroes lie inside the connected components of the density \cite{Mehta, Sze, DeiftBook}.
Our goal is also to study the asymptotic behavior of the orthogonal polynomials in the vicinity of the newborn cut.

\bigskip

{\bf Outline of the article:}\\
- In section 2 we introduce definitions and notations for orthogonal polynomials and associated quantities.\\
- In section 3 we recall classical results of random matrix theory:
the semiclassical behaviors of free energy, density, correlation functions,
orthogonal polynomials, valid away from critical points.\\
- In section 4 we study the analytical continuation of the previous
semiclassical approximations, near the ``birth of a cut'' critical point
(divergencies at critical point).\\
- In section 5 we compute the partition function with mean field theory
for eigenvalues not in the newborn cut, and derive an effective partition
function for the newborncut eigenvalues.\\
- In section 6 we use the results of section 5 to deduce the
asymptotic behaviors of correlation functions and orthogonal polynomials
in the vicinity of the critical point.\\
- Section 7 is the conclusion.\\
- In appendix, we recall Stirling's  formula and its consequences, and we recall some elliptical function basics.

\section{Setting}

Given an integer $n$ (we will later consider the limit $n\to\infty$),
 a real polynomial $V(x)$ called the potential:
\beq\label{defV}
V(x):=g_0+\sum_{k=1}^{d+1} {g_k\over k}x^k
\virg
\deg V=d+1
\eeq
of even degree and positive leading coefficient ($g_{d+1}>0$),
and a temperature $T>0$, we define the partition function:
\beq\label{defZnT}
Z_{n}(T,V):=
{1\over n!}\, \int_{{\mathbf R}^n} \D{x_1}\dots\D{x_n} \,\, \left(\Delta(x_i)\right)^2 \,\, \prod_{i=1}^n \ee{-{n\over T}
V(x_i)}
\eeq
(where $\Delta(x_i)=\prod_{i<j} (x_j-x_i)$ is the Vandermonde determinant)
and the free energy $F_n(T,V)$:
\beq\label{defF}
\ee{-{n^2\over T^2} F_n(T,V)} := {Z_n(T,V)\over H_n}
\eeq
where $H_n$ is a combinatorial normalization:
\beq\label{defHn}
H_n:= (2\pi)^{n/2}\, n^{-n^2/2}\, \ee{3/4 n^2}\, \prod_{k=0}^{n-1} k!
=2^{n/2}\,\pi^{n^2/2}\, n^{-n^2/2}\, {\ee{3/4 n^2}\over n! U_n}
\eeq
and $U_n$ is the volume of the group $U(n)/U(1)^n$.

Then we define the resolvent:
\beq\label{defW}
W_n(x,T,V) := {T\over n!\, Z_n(T,V)} \int \D{x_1}\dots\D{x_n} \,\,
{1\over x-x_1} \,\, \Delta^2(x_i) \,\, \prod_{i=1}^n \ee{-{n\over T}
V(x_i)}
\eeq
and its first moment:
\bea\label{defTrace}
\T_n(T,V) & := & {T\over n!\,Z_n(T,V)} \int \D{x_1}\dots\D{x_n} \,\,
x_1 \,\, \Delta^2(x_i) \,\, \prod_{i=1}^n \ee{-{n\over T}
V(x_i)} \cr
& = & {1\over 2i\pi} \oint x W_n(x,T,V) dx
 =  {\d F_n(T,V)\over \d g_1}
\eea
Notice that more generally for $k>0$:
\beq\label{dFdgk}
k{\d F_n(T,V)\over \d g_k}
 =  {1\over 2i\pi} \oint x^k W_n(x,T,V) dx
\eeq

\bigskip

Then, given a temperature $T_c$ and an integer $N$, we define:
\beq\label{defhn}
h_n:= {Z_{n+1}(T_c{n+1\over N},V)\over Z_n(T_c{n\over N},V)} \, ,
\eeq
and
\beq\label{defgamma}
\gamma_{n}:=  \sqrt{h_n\over h_{n-1}}
= \sqrt{Z_{n+1}(T_c{n+1\over N},V)Z_{n-1}(T_c{n-1\over N},V)
\over Z_{n}^2(T_c{n\over N},V)} \, ,
\eeq
and
\beq\label{defbetan}
\beta_n := {N\over T_c}\left( \T_{n+1}(T_c{n+1\over N},V)- \T_n(T_c{n\over N},V) \right)
=-{T_c\over N}\, {\d \ln{h_n}\over \d g_1} \, .
\eeq
Notice that $Z_n(T,V)$, $h_n$ and $\gamma_n$ are strictly positive for all $n$,
in particular they don't vanish.

We also introduce the functions \cite{Sze, Mehta, DeiftBook}:
\beq\label{deforthoPol}
\pi_n(\xi) = {Z_n(T_c{n\over N},V(x)-{T_c\over N}\ln{(\xi-x))}\over Z_n(T_c{n\over N},V)} \, ,
\virg
\psi_n(\xi) = {\pi_n(\xi)\over \sqrt{h_n}}\,\ee{-{N\over 2T}\,V(\xi)}
\eeq
which form an orthogonal family of monic polynomials ($\deg \pi_n = n$):
\beq\label{ortho}
\int \pi_n(\xi) \pi_m(\xi) \ee{-{N\over T_c} V(\xi)} d\xi = h_n \delta_{nm}\, ,
\eeq
and which satisfy the 3-terms recursion relation:
\beq\label{xrecursion}
\xi\, \pi_n(\xi) = \pi_{n+1}(\xi) + \beta_n\, \pi_n(\xi) + \gamma_n^2\, \pi_{n-1}(\xi)\, .
\eeq

And we introduce the functions:
\beq\label{deforthoPolCauchy}
\hat\pi_n(\xi) = {Z_{n+1}(T_c{n+1\over N},V(x)+{T_c\over N}\ln{(\xi-x))}\over Z_n(T_c{n\over N},V)} \, ,
\virg
\phi_n(\xi) = {\hat\pi_n(\xi)\over \sqrt{h_n}}\,\ee{{N\over 2T}\,V(\xi)}
\eeq
which are the Hilbert transforms of the $\pi_n(x)$:
\beq
\hat\pi_n(x) = \int {dx'\over x-x'}\pi_n(x')\ee{-V(x')} \, ,
\eeq
and which satisfy the same 3-terms recursion relation, with an initial term:
\beq\label{xhatrecursion}
\xi\, \hat\pi_n(\xi) = \hat\pi_{n+1}(\xi) + \beta_n\, \hat\pi_n(\xi) + \gamma_n^2\, \hat\pi_{n-1}(\xi) + \delta_{n,0} h_0 \, .
\eeq

We also define the kernel:
\beq
K_n(\xi,y) := \sum_{j=0}^{n-1} {1\over h_j}\, \pi_j(\xi)\pi_j(y)\,
\ee{-{N\over 2T_c}(V(\xi)+V(y))} \, ,
\eeq
and it is well known that all correlation functions can be expressed in term
of that kernel (Dyson's theorem \cite{thDyson, Mehta}):
\bea
&& \rho_n(x) = {1\over n}K_n(x,x) \cr
&& \rho_n(x,y) = {1\over n^2}\left(K_n(x,x)K_n(y,y) - K_n(x,y)K_n(y,x) \right) \cr
&& \dots
\eea
and that we have the Christoffel-Darboux theorem:
\beq
K_n(x,y) = {1\over h_{n-1}}\,
{\pi_n(x)\pi_{n-1}(y)-\pi_{n-1}(x)\pi_n(y)\over x-y}\,
\ee{-{N\over 2T_c}(V(x)+V(y))} \, .
\eeq

\bigskip
Our goal is to study $\gamma_n$, $\beta_n$, $\pi_n$, $\hat{\pi}_n$,
in the vicinity of $N\to\infty$, and $|n-N|\ll N$,
and with $T_c$ chosen such that we are at a special
critical point described below.

For the moment, let us study the large $N$-limits away from the critical point.

\section{Classical limits}
\label{sectionclassical}

It is well known that in the semiclassical limit
$N\to\infty$, and $|n-N|\ll N$, if $T\neq T_c$,
the free energy has a large $n$ limit \cite{dkmvz, EML, BDE, Joh, ZJDFG}:
\beq\label{largenF}
F_n(T,V) \longrightarrow F(T,V) +O(1/n^2)
\eeq
and so has the resolvent $W_n(x,T,V)$:
\beq\label{largenW}
W_n(x,T,V) \longrightarrow W(x,T,V) +O(1/n)
\eeq

\subsection{The large $n$ resolvent}

It is well known that, if the potential is such that
$V'(x)$ is a rational fraction, with its poles outside the cuts
(that assumption will become clear below),
the large $n$ resolvent $W(x,T,V)$ can be written as the solution of an
hyperelliptical equation \cite{ZJDFG, BIPZ}:
\beq
W(x,T,V) = {1\over 2}\left(
V'(x) - M(x,T,V)\sqrt{\sigma(x,T,V)}
\right)
\eeq
where $\sigma$ is a monic even
degree ($2s\geq 2$) polynomial with distinct simple zeroes only:
\beq
\sigma(x) = \prod_{i=1}^{s} (x-a_i)(x-b_i)
\virg
\dots<a_i<b_i<a_{i+1}<\dots
\eeq
whose zeroes are called the endpoints, and $\bigcup_{i=1}^s [a_i,b_i]$
is called the support,
and $M$ is a rational function with the same poles as $V'$.
If one assumes that $s$ and $\sigma$ are known, $M(x,T,V)$ is determined by
the condition that $W(x,T,V)$ is finite (in the physical sheet)
when $x\to \infty$ and when $x$ approaches the poles of $V'(x)$.

\medskip

The large $n$ limit of the density of eigenvalues is then:
\beq
\rho(x,T,V) = {1\over 2\pi T} M(x,T,V)\sqrt{-\sigma(x,T,V)}
\virg
x\in \bigcup_{i=1}^s [a_i,b_i]
\eeq

We also define the effective potential:
\beq
V_{\rm eff}(x,T,V) := V(x) -2T\ln{x} - 2 \int_{\infty}^x
\left( W(x,T,V)-{T\over x}\right) dx
\eeq
Notice that its derivative is $V'(x)-2W(x,T,V)=M(x,T,V)\sqrt{\sigma(x,T,V)}$.

So far, we have not explained how to determine $s$ and the polynomial $\sigma$.
If one assumes that $s$ is known, $\sigma(x,T,V)$ is determined by the
conditions:
\beq
\displaystyle
\left\{
\begin{array}{l}
\displaystyle W(x,T,V) \mathop\sim_{x\to\infty} {T\over x} + O(1/x^2) \cr
\displaystyle {\rm if\,}\, s>1,\quad
\forall i=1,\dots,s-1,  \,\,\,
V_{\rm eff}(b_i) = V_{\rm eff}(a_{i+1})
\end{array}
\right.
\eeq

The large $n$ free energy is then given by:
\beq
F(T,V) = {1\over 4i\pi}\oint W(x,T,V) V(x) dx
+ {1\over 2}T V_{\rm eff}(b_s)
\eeq
where the integration contour is a counter clockwise circle around $\infty$.

The number of endpoints $s=s(T,V)$, (we have $1\leq s\leq d$) is determined
by the condition that the free energy is minimum
(one can prove that $s(T,V)\leq (d+1)/2$) \cite{Matrixsurf}.

\subsection{Derivatives with respect to $T$}
\label{sectionderivativesT}

Let us introduce:
\beq\label{defOmega}
\Omega(x,T,V) :={\d W(x,T,V)\over \d T} = {Q_\Omega(x,T,V)\over \sqrt{\sigma(x,T,V)}}
\eeq
where $Q_\Omega(x,T,V)$ is a monic polynomial of degree $s-1$,
determined by the conditions:
\beq
\displaystyle {\rm if\,}\, s>1,\quad
\forall i=1,\dots,s-1,  \,\,\,
\int_{b_i}^{a_{i+1}} {Q_\Omega(x,T,V)\over \sqrt{\sigma(x,T,V)}}\, dx = 0
\eeq
In algebraic geometry, $\Omega$ is called ''normalized abelian differential of the third kind'' \cite{Fay, Farkas}.

We introduce the multivalued function $\L(x,T,V)$:
\beq\label{defLambda}
\L(x,T,V) := \exp{\left(\int_{b_s}^x \Omega(x',T,V)\, dx'\right)}
\eeq
and
\beq\label{defgammalim}
\gamma(T,V):= \mathop{{\rm lim}}_{x\to\infty} \, {x\over \L(x,T,V)}
\eeq

Then we have the following derivatives:
\beq\label{dFdT}
{\d F\over \d T} = V_{\rm eff}(b_s)
\eeq

\beq\label{ddFdTdT}
{\d^2 F\over \d T^2} = -2\ln\gamma
\eeq

\beq\label{dVeffdT}
{\d V_{\rm eff}(x)\over \d T} = -2\ln{(\gamma\L(x))}
\eeq

\beq\label{dcalTdT}
{\d {\cal T}\over \d T} = {1\over 2i\pi}\,\oint x\, \Omega(x)\, \D x
\eeq

Notice that:
\beq
V_{\rm eff}(b_s) = {1\over 2i\pi}\oint \Omega V  - 2 T \ln\gamma
\eeq

\subsection{Poles of the potential}
\label{sectionderivativesr}

Assume that $V'(x)$ has a simple pole at $x=\xi$, with residue $r$
(it may have other poles too),
then we define the function:
\beq\label{defHdWdh}
H(x,\xi,T,V):={\d W(x,T,V)\over \d r} = {1\over 2\sqrt{\sigma(x)}}
\left(
{\sqrt{\sigma(x)}-\sqrt{\sigma(\xi)}\over x-\xi}
-Q_H(x,\xi)
\right)
\eeq
where $Q_H(x,\xi)$ is a monic polynomial in $x$, of degree $s-1$,
determined by the conditions:
\beq
\displaystyle {\rm if\,}\, s>1,\quad
\forall i=1,\dots,s-1,  \,\,\,
\int_{b_i}^{a_{i+1}} {Q_H(x,\xi,T,V)+{\sqrt{\sigma(\xi)}\over x-\xi}
\over \sqrt{\sigma(x,T,V)}}\, dx = 0
\eeq
We also define its (multivalued) primitive:
\beq\label{defprimeform}
\ln{E(x,\xi)}:=\int_{\infty}^x H(x',\xi) \, d x'
\eeq
Notice that it is finite near the endpoints and near $x=\xi$.
In algebraic geometry, $(x-\xi)/E(x,\xi)$ is related to the ''prime form'' \cite{Fay, Farkas}.

Then we have:
\beq\label{dVeffdr}
{\d V_{\rm eff}(x)\over \d r} = \ln{(x-\xi)}-2\ln{E(x,\xi)}
\eeq

\beq\label{dcalTdr}
{\d {\cal T}\over \d T} = {1\over 2i\pi}\,\oint x\, H(x,\xi)\, \D x
\eeq

\beq
{\d F\over \d r} = {1\over 2}\left.(V(x)-V_{\rm eff}(x))\right|_{x=\xi}
\eeq
\beq
{\d^2 F\over \d r \d T} = \ln{(\gamma\L(\xi))} = \ln{(\xi-b_s)}-2 \ln{E(b_s,\xi)}
\eeq

If $V'(x)$  has simple poles at $x=\xi_1$ with residue $r_1$ and at
$x=\xi_2$ with residue $r_2$, we have:
\beq
{\d^2 F\over \d r_1 \d r_2} = \ln{E(\xi_1,\xi_2)}
\eeq
and thus it is clear that $\ln{E}$ has some symmetry properties: $\ln{E(x,y)}=\ln{E(y,x)}$.

\subsection{$1$-cut case}
\label{section1cutparam}

If $W(x,T,V)$ has one cut $[a(T,V),b(T,V)]$ with $a<b$,
we use the Joukowski's parameterization:
\beq
x = {b+a\over 2} + {b-a\over 2}\cosh\phi
\eeq
i.e.
\beq
\sqrt{\sigma(x)} = {b-a\over 2}\sinh\phi
\eeq

We have:
\beq\label{Omegaonecut}
\Omega(x) = {1\over \sqrt{(x-a)(x-b)}}={\d \phi\over \d x}
\virg
\L(x) = \ee{\phi(x)}
\virg
\gamma={b-a\over 4}
\eeq

\beq
H(x,\xi) = {\d\phi(x)\over \d x}\,\,{1\over \ee{\phi(x)+\phi(\xi)}-1}
\virg
E(x,\xi) = 1-\ee{-(\phi(x)+\phi(\xi))} = {x-\xi\over \L(x)-\L(\xi)}
\eeq

\beq\label{dcalTdTronecut}
{\d {\cal T}\over \d T} = {a+b\over 2}
\virg
{\d {\cal T}\over \d r} = {\gamma\over \L(\xi)}
\eeq

\medskip

Then, it is well known \cite{} that we have the large $n,N$ asymptotics (in the regime $n/N=$finite):
\beq\label{asympgammabetaonecut}
\gamma_n \sim {b(T_c{n\over N})-a(T_c{n\over N})\over 4}
\virg
\beta_n \sim {b(T_c{n\over N})+a(T_c{n\over N})\over 2}
\eeq

\subsection{$2$-cut case}
\label{section2cutparam}

If $W(x,T,V)$ has two cuts $[a(T,V),b(T,V)]\cup[c(T,V),d(T,V)]$
with $a<b<c<d$,
Let $m$ be their biratio:
\beq\label{biratio}
m={(b-a)(d-c)\over (c-a)(d-b)}
\eeq

We parameterize:
\beq\label{paramtwocuts}
x(u) = d-{d-a\over 1+{b-a\over d-b}\sn^2(u,m)}
\eeq
where $\sn$ is tha elliptical sine function (see appendix I, or for instance \cite{elliptical}), i.e., by definition:
\beq
u(x):=-{i\over 2}\,\sqrt{(d-b)(c-a)}\,\, \int^x_a {\D y\over \sqrt{\sigma(y)}}
= \int_0^{\sqrt{{d-b\over b-a}{x-a\over d-x}}}{dy\over \sqrt{(1-y^2)(1-my^2)}}
\eeq
We have:
\bea
u(a)=0 ,& u(b)=K(m), \cr
u(c)=K(m)+iK'(m) ,& u(d)=iK'(m).
\eea
We have:
\beq
\sqrt{\sigma(x)} = -i (d-a)(b-a)\sqrt{c-a\over d-b}\,\,{\sn(u,m)\,\cn(u,m)\,\dn(u,m)\over (1+{b-a\over d-b}\sn^2(u,m))^2}.
\eeq

Let us define $u_\infty$ such that:
\beq\label{uinfty}
u_\infty := i\int_0^{\sqrt{{d-b\over b-a}}}{dy\over \sqrt{(1+y^2)(1+my^2)}},
\eeq
i.e.
\beq\label{snuinfty}
\sn(u_\infty,m) = i\sqrt{d-b\over b-a},
\,\, , \,\, \cn(u_\infty,m) = \sqrt{d-a\over b-a},
\,\, , \,\, \dn(u_\infty,m) = \sqrt{(d-a)\over (c-a)}.
\eeq

Then we define $x_0$:
\beq\label{x0twocuts}
x_0 = d
+ i\sqrt{(c-a)(d-b)} \left(E(u_\infty,m)-\left(1-{E'(m)\over K'(m)}\right)u_\infty \right).
\eeq
It satisfies:
\beq
\int_b^c {x-x_0\over \sqrt{(x-a)(x-b)(x-c)(x-d)}}dx =0,
\eeq
thus we have:
\beq\label{Omegatwocuts}
\Omega(x) ={x-x_0\over \sqrt{(x-a)(x-b)(x-c)(x-d)}}
\eeq

\beq
\Lambda(x) =
\ee{\pi {u(x) u_\infty\over K K'}}\,{\theta_1((u(x)+u_\infty)/2K)\over \theta_1((u(x)-u_\infty)/2K)}
\eeq
\bea\label{gamma2cuts}
\gamma
&=& {i\over 4K} \sqrt{(d-b)(c-a)}\,\ee{-\pi{u_\infty^2\over K K'}}\,{\theta_1'(0,\tau)\over \theta_1(u_\infty/K,\tau)}
\eea
\beq
E(x,\xi)= {\theta_1(u(x)+u(\xi))\,\theta_1(2u_\infty)\over \theta_1(u_\infty+u(\xi))\,\theta_1(u_\infty+u(x))}
\eeq

\beq\label{dcalTdTtwocuts}
{\d {\cal T}\over \d T} = {a+b+c+d\over 2}-x_0
\eeq

we have the asymptotics \cite{dkmvz,BDE}:
\beq\label{boundsgamma2cuts}
{d(T_c{n\over N})-a(T_c{n\over N})-c(T_c{n\over N})+b(T_c{n\over N})\over 4}
\leq \gamma_n
\leq {d(T_c{n\over N})-a(T_c{n\over N})+c(T_c{n\over N})-b(T_c{n\over N})\over 4}
\eeq
\beq\label{boundsbeta2cuts}
{d(T_c{n\over N})+a(T_c{n\over N})-c(T_c{n\over N})+b(T_c{n\over N})\over 2}
\leq \beta_n
\leq {d(T_c{n\over N})+a(T_c{n\over N})+c(T_c{n\over N})-b(T_c{n\over N})\over 2}
\eeq

\bigskip
Therefore, we shall now study $W(x,T,V)$ in different regimes.

\section{The Birth of a cut critical point}

Let us choose the potential $V$ and the temperature $T_c$ such that:
\begin{itemize}
\item for $T<T_c$ we are in a one-cut case,
\beq\label{WbeforeTc}
W(x,T) = {1\over 2} \left( V'(x) - M_-(x,T) \sqrt{(x-a)(x-b)} \right)
\eeq
with $a(T)<b(T)$ and
\bea
  M_-(x,T_c)=(x-e)^{2\nu-1}\,Q(x)\, ,
\eea
where $\nu\geq 1$ is an integer, and $Q(x)$ is a real polynomial whose properties are described bellow.

\item for $T>T_c$ we are in a two-cuts case,
\beq\label{WafterTc}
W(x,T) = {1\over 2} \left( V'(x) - M_+(x,T) \sqrt{(x-a)(x-b)(x-c)(x-d)} \right)
\eeq
with $a(T)<b(T)<c(T)<d(T)$ and
\bea
c(T_c)=d(T_c)=e
&,& M_+(x,T_c)=(x-e)^{2\nu-2}\,Q(x)\, ,
\eea

\item at $T=T_c$ one cut has vanishing size $c(T_c)=d(T_c)$.
With no loss of generality, we can assume that:
\beq
a(T_c)=-2 \virg b(T_c)=2,
\eeq
and we write:
\beq\label{defe}
e(T_c)=2\cosh{\phi_e}=c(T_c)=d(T_c)
\eeq
\end{itemize}

\medskip

The polynomial $Q(x)$ must have the following properties:
\begin{itemize}
\item $\deg Q=d-2\nu$ with $d$ odd and $d>2\nu$,
\item The leading coefficient of $Q$ is positive,
\item $Q$ has an odd number of zeroes in $]2,e[$,
\item $Q(x)<0$ in $[-2,2]$,
\item $Q(e)> 0$,
\item
\beq
\forall x<-2,
\quad
\int_x^{-2} Q(x)(x-e)^{2\nu-1}\sqrt{x^2-4}\, dx \, >0
\eeq
\item
\beq
\forall x>2, \,\, x\neq e,
\quad
\int_{2}^x Q(x)(x-e)^{2\nu-1}\sqrt{x^2-4}\, dx \, >0
\eeq
\item
\beq\label{poteffvanishate}
\int_{2}^e Q(x)(x-e)^{2\nu-1}\sqrt{x^2-4}\, dx =0
\eeq
\item
\beq\label{VpolQ}
V'(x) = \Pol_{x\to\infty} \left( (x-e)^{2\nu-1}Q(x)\sqrt{x^2-4}\right)
\eeq
\item
\beq\label{TcresQ}
T_c = {1\over 2}\Res_{\infty}\, (x-e)^{2\nu-1}Q(x)\sqrt{x^2-4} \,dx
\eeq
\end{itemize}

\bigskip
{\noindent \bf Remark:}
notice that for all $\nu\geq 1$, it is possible to find a potential $V(x)$ and a temperature $T_c$
with such properties.
Indeed, choose $e$ and $Q(x)$ with the above properties and determine $V'(x)$ and $T_c$ by \ref{VpolQ} and \ref{TcresQ}.
Notice also that it is always possible to find a polynomial $Q(x)$ which satisfies the above mentioned conditions,
indeed consider any real $e>2$, and any real polynomial $\td{Q}(x)$, of even degree $d-2\nu-1$,
with positive leading coefficient, and with no real zero,
then set:
\beq
\td{e} = {\int_{2}^e x\,\td{Q}(x)\,(x-e)^{2\nu-1}\sqrt{x^2-4}\, dx\over \int_{2}^e \td{Q}(x)\,(x-e)^{2\nu-1}\sqrt{x^2-4}\, dx}
\eeq
clearly, $\td{e}\in ]2,e[$, and then set:
\beq
Q(x)=(x-\td{e})\td{Q}(x)
\eeq

In particular, one may choose $d=2\nu+1$ and $\td{Q}=1$.

\subsection{Example $\nu=1$}

Let $e>2$ be fixed. We write $e=2\cosh{\phi_e}$.

We consider the following quartic potential:

\beq
V'(x) = \left( x^3 - (e+\td{e}) x^2 + (e\td{e}-2) x + 2 (e+\td{e}) \right)
\virg T_c=1+e\td{e}
\eeq
where $\td{e}$ is given by $\int_2^e (x-e)(x-\td{e}) \sqrt{x^2-4} = 0$, i.e. :
\beq
\td{e}=2 {\phi_e \cosh\phi_e - {1\over 3}\sinh\phi_e(2+\cosh^2\phi_e)\over
{1\over 3}\sinh\phi_e \cosh\phi_e(5-2\cosh^2\phi_e) - \phi_e}
\eeq

\subsection{At the critical point $T=T_c$:}

At $T=T_c$, both formula \ref{WbeforeTc} and \ref{WafterTc} reduce to:
\beq
W(x,T_c) = {1\over 2} \left( V'(x) - (x-e)^{2\nu-1}Q(x) \sqrt{x^2-4} \right)
\eeq
which would correspond to an average large $N$ eigenvalue density in $[-2,2]$:
\beq
\rho(x) =  {1\over 2\pi T_c} (x-e)^{2\nu-1}Q(x) \sqrt{4-x^2}
\eeq
and one would have:
\beq
\gamma_N \sim 1
\virg
\beta_N \sim 0
\eeq

However, this is wrong, because the semiclassical asymptotics \ref{largenW} are valid
only if $T\neq T_c$, they  break down at $T=T_c$.
It is the purpose of section \ref{secasympu}, to determine the asymptotic behavior of $\gamma_n$ and $\beta_n$ near $n=N$ and $T=T_c$.
For the moment, let us consider the limits of \ref{WbeforeTc} and \ref{WafterTc} near $T_c$.

\subsection{Variations near the critical point, $T<T_c$ (one-cut)}

Let us consider the limit of \ref{WbeforeTc} near $T_c$.
Write $T= T_c+t$, and $t<0$, and:
\beq
W(x,T) = {1\over 2} \left( V'(x) - M_-(x,T) \sqrt{(x-a)(x-b)} \right)
\eeq
At $T=T_c$ we have
\bea
a(T_c)=-2 \virg b(T_c)=2 \virg  M_-(x,T_c)=(x-e)^{2\nu-1}\,Q(x)\, ,
\eea

Then, make use of formula  \ref{defOmega} and \ref{Omegaonecut}, i.e.
\beq\label{dWdTonecut}
-{1\over 2}\,{\d M_-(x,T)\over \d T}
+{1\over 4}\,M_-(x,T)\,{{\d a\over \d T}\over (x-a)}\,+{1\over 4}\,M_-(x,T)\,{{\d b\over \d T}\over (x-b)}\,
={1\over (x-a)(x-b)}
\eeq
matching the pole at $x=a$ gives:
\beq
{\d a\over \d T} ={4\over (a-b)\,M_-(a,T)}
\sim -{1\over (a-e)^{2\nu-1}Q(a)}
\eeq
which is finite at $T=T_c$, thus, we find that to first order in $t$:
\beq
a \sim -2 + {t\over (2+e)^{2\nu-1}Q(-2)}
\virg
b \sim 2 - {t\over (e-2)^{2\nu-1}Q(2)}
\eeq
(notice that $Q(-2)<0$ and $Q(2)<0$).
Relation \ref{Omegaonecut} implies:
\beq\label{gammaonecutval}
\gamma \sim 1 + O(t)
\eeq

Then, \ref{dWdTonecut} reduces to:
\beq
{1\over 2}\,{\d M_-(x,T)\over \d T}
={M_-(x,T)-M_-(a,T)\over M_-(a,T)(a-b)(x-a)}\,
+{M_-(x,T)-M_-(b,T)\over M_-(b,T)(b-a)(x-b)}\,
\eeq
which is finite at $T=T_c$, thus one gets the asymptotics of $M_-$:
\bea
2\,{\d M_-(x,T)\over \d T}
&=& -{(x-e)^{2\nu-1}(Q(x)-Q(a))+((x-e)^{2\nu-1}-(a-e)^{2\nu-1})Q(a)\over (a-e)^{2\nu-1}\,Q(a)\,(x-a)}\, \cr
&& +{(x-e)^{2\nu-1}(Q(x)-Q(b))+((x-e)^{2\nu-1}-(b-e)^{2\nu-1})Q(b)\over (b-e)^{2\nu-1}\,Q(b)\,(x-b)}\, \cr
&=& -{(x-e)^{2\nu-1}\over (a-e)^{2\nu-1}\,Q(a)}\,{Q(x)-Q(a)\over x-a}
 -{(x-e)^{2\nu-1}-(a-e)^{2\nu-1}\over (x-a)\,(a-e)^{2\nu-1}}\, \cr
&& +{(x-e)^{2\nu-1}\over (b-e)^{2\nu-1}\,Q(b)}\,{Q(x)-Q(b)\over x-b}
 +{(x-e)^{2\nu-1}-(b-e)^{2\nu-1}\over (x-b)\,(b-e)^{2\nu-1}}\, \cr
\eea

in particular in the vicinity of $x=e$ one has:
\bea
M_-(x,T)&\sim& (x-e)^{2\nu-1}Q(x) \cr
&& +{t\over 2}\left[
\sum_{k=0}^{2\nu-2} (x-e)^k ((2-e)^{-k-1}-(-2-e)^{-k-1})
+O(x-e)^{2\nu-1}\right] \cr
&&
\eea
Note that the zeroes of $M_-(x,T)$ in the vicinity of $e$, are the $2\nu-1^{\rm th}$ roots of unity:
\beq
M_-(x,T_c+t)=0 \quad \leftrightarrow\quad
x=e+\left({2t\over Q(e)(e^2-4)}\right)^{1/2\nu-1} + O(t^{2/2\nu-1})
\eeq

Using \ref{asympgammabetaonecut}, we get:
\bea
&& \gamma_n \sim 1-{t\over 4} \left({1\over (e-2)^{2\nu-1}Q(2)}+{1\over (e+2)^{2\nu-1}Q(-2)}\right) \cr
&& \beta_n \sim -{t\over 2} \left({1\over (e-2)^{2\nu-1}Q(2)}-{1\over (e+2)^{2\nu-1}Q(-2)}\right)   \cr
&& {\rm where} \quad n=N(1+t/T_c)
\eea

\subsection{Variations near the critical point, $T>T_c$ (two cuts)}

Let us consider the limit of \ref{WafterTc} near $T_c$.
Write $T= T_c+t$, and $t>0$,and
\beq
W(x,T) = {1\over 2} \left( V'(x) - M_+(x,T) \sqrt{(x-a)(x-b)(x-c)(x-d)} \right)
\eeq
At $T=T_c$ we have
\bea
a(T_c)=-2 &,& b(T_c)=2 \cr
c(T_c)=d(T_c)=e=2\cosh{	\phi_e}
&,& M_+(x,T_c)=(x-e)^{2\nu-2}\,Q(x)\, ,
\eea
and at $T>T_c$, $a+2$, $b-2$, $c-e$, $d-e$, and $M_+(x)-(x-e)^{2\nu-2}Q(x)$ are small.
In particular, we write:
\beq
M_+(x,T) = H(x,T) Q(x,T)
\eeq
where $H(x,T)$ is a monic polynomial of degree $2\nu-2$ which contains all the roots of $M_+$ close to $e$, and $Q(x,T)$ is the remaining part.
In other words, $H(x,T)-(x-e)^{2\nu-2}$ is small and $Q(x,T)-Q(x)$ is small in the small $t$ limit.

We use the notations of section \ref{section2cutparam}.
The biratio \ref{biratio} is thus:
\beq
m={(b-a)(d-c)\over (c-a)(d-b)}
\sim {4\over e^2-4}\,(d-c)
\sim {d-c\over \sinh^2{\phi_e}}
\eeq
we see that we have to consider the limit $m\to 0$.
In that limit \ref{uinfty} becomes
\beq
u_\infty = i\int_0^{\sqrt{{d-b\over b-a}}}{dy\over \sqrt{(1+y^2)(1+my^2)}} \sim      i\int_0^{\sqrt{{e-2\over 4}}} {dy\over \sqrt{1+y^2}} =        i{\phi_e\over 2}
\eeq

\bea
E(u_\infty)-u_\infty
&=& im\, \int_0^{\sqrt{{d-b\over b-a}}} {y^2\,dy\over \sqrt{(1+y^2)(1+my^2)}} \cr
&\sim& im\, \int_0^{\sqrt{{e-2\over 4}}} {y^2\,dy\over \sqrt{(1+y^2)}} = im\,{\sinh\phi_e-\phi_e \over 4} \cr
\eea
And, as can be found in any handbook of classical functions \cite{}, we have the small $m$ behavior:
\beq
{E'(m)\over K'(m)} \sim -{2\over \ln{m}}
\eeq
Since for small $m$ one has $\left|{1\over \ln{m}}\right|\gg m$, \ref{x0twocuts} becomes:
\beq\label{asympx0}
\delta x_0:=x_0-d  \sim  {2\phi_e \sinh{\phi_e}\,
\over \ln{m}}
\eeq
Notice that:
\beq
x_0-c=x_0-d+d-c \sim x_0-d+m \sinh^2{\phi_e}   \sim x_0-d \sim \delta x_0
\eeq

From \ref{defOmega} and \ref{Omegatwocuts} we have:
\beq\label{dMsigmaOmega}
-{1\over 2}\,{\d M_+(x,T)\over \d T}
-{1\over 4}\,{M_+(x,T)\over \sigma(x,T)}\,{\d \sigma(x,T)\over \d T}
={x-x_0\over \sigma(x,T)}
\eeq
Matching the pole at $x=c$ gives:
\bea\label{dcdteq}
{\d c\over \d t}
 =  {-4(d-c+\delta x_0)\over (c-a)(c-b)(c-d)M_+(c,T_c+t)}
 \sim  {-4 \delta x_0\over (e^2-4)Q(e)}\, {1\over (c-d)\,H(c)}
\eea
and matching the pole at $x=d$ gives:
\bea\label{dddteq}
{\d d\over \d t}
 =  {-4\delta x_0\over (d-a)(d-b)(d-c)M_+(d)}
 \sim  -{4\delta x_0\over (e^2-4)Q(e)}\, {1\over (d-c)\,H(d)}
\eea
and matching the poles close to $e$ gives:
\bea\label{dHdteq}
{\d H(x)\over \d t} &\sim& -{2\delta x_0\over (e^2-4)Q(e)}\,{1\over d-c}\,\left({H(x)-H(d)\over (x-d)H(d)}-{H(x)-H(c)\over (x-c)H(c)}\right)
\eea

The following guess solves the 3 equations \ref{dcdteq}, \ref{dddteq}, \ref{dHdteq} to small $t$ leading order:

\beq\label{ansatzcd}
c \sim e- 2\zeta\, \left(-{t\over \ln{t}}\right)^{1\over 2\nu}
\virg
d \sim e+2\zeta\, \left(-{t\over \ln{t}}\right)^{1\over 2\nu}
\eeq
\beq\label{ansatzH}
H(x) \sim \left(-{t\over \ln{t}}\right)^{-1+{1 \over \nu}} \,\, G\left((x-e)\left(-{t\over \ln{t}}\right)^{-1\over 2\nu}\right)
\eeq
where $\zeta$ is a positive real number, and $G$ is a degree $2\nu-2$ even monic polynomial, which will be determined below.

For later convenience, we also define the following positive constant:
\beq
C:={4\nu^2\phi_e\over \sinh{\phi_e}\,Q(e)} >0
\eeq

Using ansatz \ref{ansatzcd} and \ref{asympx0}, we have in that limit:
\beq
\delta x_0  \sim  {4\nu \phi_e \sinh{\phi_e}\, \over \ln{t}}
\eeq

Then, inserting \ref{ansatzcd} and \ref{ansatzH} into \ref{dcdteq} and \ref{dddteq}, we get:
\beq\label{relzetaG1}
4\zeta^2 =  {C\over  G(-2\zeta)} =  {C\over  G(2\zeta)}
\eeq

Then, setting $x=e+\xi \left(-{t\over \ln{t}}\right)^{1\over 2\nu}$, and inserting \ref{ansatzH} into \ref{dHdteq} we get the following equation for $G$:
\beq
{(2\nu-2) \,G(\xi)-\xi\,G'(\xi)}
 =   {C\over 4\zeta}\,\,\left({G(\xi)-G(2\zeta)\over (\xi-2\zeta)G(2\zeta)}-{G(\xi)-G(-2\zeta)\over (\xi+2\zeta)G(-2\zeta)}\right)
\eeq
which using \ref{relzetaG1} becomes:
\beq\label{eqdifGxi}
{(2\nu-2) \,G(\xi)-\xi\,G'(\xi)} =  {4\zeta^2\over \xi^2-4\zeta^2}\,\,(G(\xi)-G(2\zeta))
\eeq
the solution of which is:
\beq\label{Gxival}
G(\xi)
=
\sum_{k=0}^{\nu-1} {2k!\over k! k!} \,\zeta^{2k}\,  \xi^{2(\nu-1-k)}
={\rm Pol}\,{\xi^{2\nu-1}\over \sqrt{\xi^2-4\zeta^2}}
\eeq
or:
\beq
G(2\zeta \cosh\psi)= \zeta^{2\nu-2}\,\sum_{j=0}^{\nu-1}\, \pmatrix{2\nu-1\cr \nu+j}\,{\sinh{(2j+1)\psi}\over \sinh\psi}
\eeq

In particular,
\beq
{G(2\zeta)\over \zeta^{2\nu-2}}
={1\over 2}\,{2\nu!\over   \nu-1! \nu!}
= {C\over 4 \zeta^{2\nu}}
\eeq
i.e. the parameter $\zeta$ is determined by:
\beq\label{zetaC}
\zeta
= \left({C\over 2}\,{\nu!\, \nu-1! \over 2\nu !}\right)^{1\over 2\nu}
= \left({2\nu^2\phi_e\over \sinh{\phi_e}\,Q(e)}\,{\nu!\, \nu-1! \over 2\nu !}\right)^{1\over 2\nu}
\eeq

\bigskip

In that scaling regime, we have:
\beq
m \sim {4\zeta\over \sinh^2{\phi_e}}\, \left(-{t\over \ln{t}}\right)^{1\over 2\nu}
\eeq
i.e. this corresponds to a torus of modulus
\beq
\tau=i\,{K'\over K}\sim {-i\over \pi} \ln{m}
\sim {-i\over 2\nu\pi} \ln{t}
\eeq
and, using \ref{gamma2cuts}:
\bea\label{gammatwocutsval}
\gamma
&=& {i\over 4K} \sqrt{(d-b)(c-a)}\,\ee{-\pi{u_\infty^2\over K K'}}\,{\theta_1'(0,\tau)\over \theta_1(u_\infty/K,\tau)} \cr
&\sim&\ee{-\pi{u_\infty^2\over K K'}} \cr
&\sim& 1-{\phi_e^2\over \ln{m}} \cr
&\sim& 1-{2\nu\phi_e^2\over \ln{t}} \cr
\eea

we also find that the filling fraction in the $[c,d]$ cut is of order:
\bea
\epsilon &=&{1\over 2\pi T_c}\int_c^d \rho(x)\, \D{x} \cr
&\sim& -{t\over \ln{t}} \,\,{\sinh{\phi_e}Q(e)\over \pi T_c} \, \int_{-2\zeta}^{2\zeta} G(\xi) \sqrt{4\zeta^2-\xi^2}\, \D\xi \cr
&\sim& -i{t\over \ln{t}} \,\,{\sinh{\phi_e}Q(e)\over \pi T_c} \, \int_{-2\zeta}^{2\zeta} \xi^{2\nu-1}\,{G(\xi)\sqrt{4\zeta^2-\xi^2}\over \xi^{2\nu-1}}\,  \D\xi  \cr
\eea
then, integrating by parts and using \ref{eqdifGxi}, we find:
\bea
\epsilon
&\sim& i{t\over T_c\,\ln{t}} \,\,{\sinh{\phi_e}Q(e)\over 2\nu\pi} \, \int_{-2\zeta}^{2\zeta} \xi^{2\nu}\, {4\zeta^2\,\xi^{-2\nu}\,\over \sqrt{\xi^2-4\zeta^2}}\,\,G(2\zeta)  \,\D\xi   \cr
&\sim& i{t\over T_c\,\ln{t}} \,\,{\sinh{\phi_e}Q(e)\over 2\nu\pi } \,4\zeta^2\,\,G(2\zeta) \, \int_{-2\zeta}^{2\zeta} {\D\xi\over \sqrt{\xi^2-4\zeta^2}}    \cr
&\sim& -{t\over T_c\,\ln{t}} \,\,{\sinh{\phi_e}Q(e)\over  2\nu} \,C     \cr
\eea
i.e.
\beq
\epsilon(T_c+t)\sim -{t\over T_c\,\ln{t}} \,\,2\nu\phi_e
\eeq
This means that for $n=N(1+t/T_c)$, the average number of eigenvalues located near $e$ is:
\beq
k \sim {n-N\over \ln{N}} \,\,2\nu\phi_e
\eeq
i.e. the eigenvalues start to explore the potential well near $e$ when $n-N\sim \ln{N}$.

According to \ref{boundsgamma2cuts} and \ref{boundsbeta2cuts}, the coefficients $\gamma_n$ and $\beta_n$ vary between:
\beq
1+\zeta \left({-t\over \ln{t}}\right)^{1\over 2\nu}
\leq \gamma_n \leq \cosh{\phi_e}
\eeq
\beq
2\zeta \left({-t\over \ln{t}}\right)^{1\over 2\nu} \leq \beta_n
\leq e-2
\eeq
$$
\quad {\rm where} \quad
n=N(1+t/T_c)
$$

The transition takes place on a scale of order $p\sim \ln{N}$.

\subsection{Order of the transition}

\begin{figure}[bth]
\hrule\hbox{\vrule\kern8pt
\vbox{\kern8pt \vbox{
\begin{center}
{\mbox{\epsfxsize=10.truecm\epsfbox{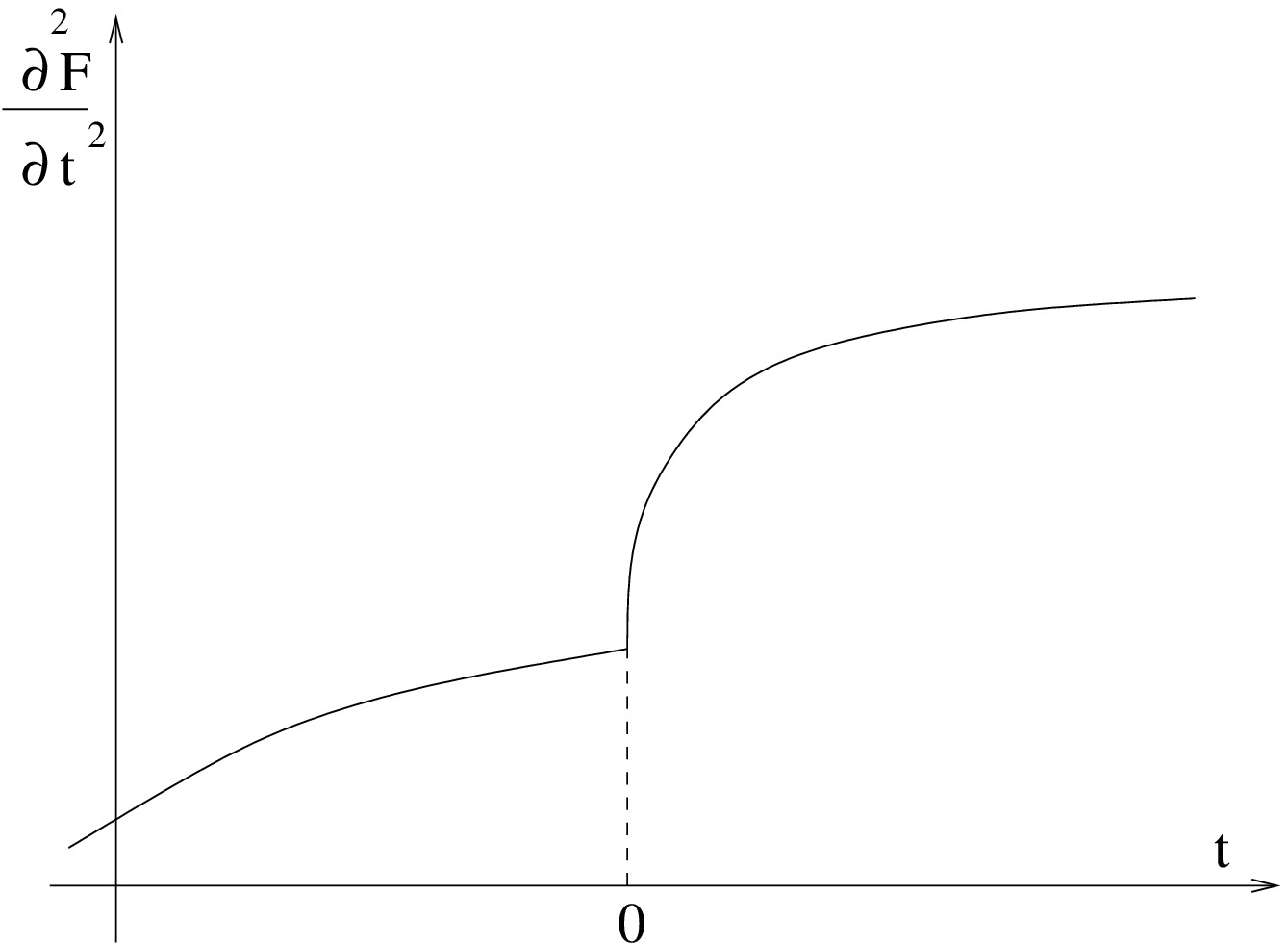}}}
\end{center}
\caption{Behavior of the second derivative of the classical free energy with respect to $t$ near the transition.\label{figtransitionclas}}
}\kern8pt}
\kern8pt\vrule}\hrule
\end{figure}

From \ref{ddFdTdT} and \ref{gammaonecutval} we have below $T_c$, i.e. for $t<0$:
\beq
{\d^2 F\over \d t^2}(T_c+t) =-2\ln\gamma \sim {t\over 2} \left({1\over (e-2)^{2\nu-1}Q(2)}+{1\over (e+2)^{2\nu-1}Q(-2)}\right)
\eeq
and above $T_c$, i.e. for $t>0$, we have from \ref{gammatwocutsval}:
\beq
{\d^2 F\over \d t^2}(T_c+t) =-2\ln\gamma \sim { 4\nu \phi_e^2\over \ln{t}}
\eeq
The second derivative of the free energy is continuous, but the third derivative is not.
therefore we have a third order transition, with logarithmic divergency, cf fig.\ref{figtransitionclas}.

\section{Mean field asymptotics for the partition function}
\label{secasympu}

We compute the partition function for a potential:
\beq
V_{r_0}(x) = V(x)+r_0\ln{(x_0-x)}
\eeq
where $r_0$ is assumed of order $1/N$:
\beq
r_0 = - a {T_c\over N}
\eeq

\subsection{Mean Field theory}

We use the same idea as in \cite{BDE}.
We split the 2-cuts integral into 1-cut integrals.
Let us say that there are $k$
eigenvalues in the new cut (near $e$), and $n-k$ in the old cut $[a,b]$:
\bea
&& n!\, Z_{n}({n\over N}T_c,V_{r_0}) \cr
&\sim& \sum_{k=0}^{n} \pmatrix{n \cr k}
\int_{x_i>\td{e}} \D{x_1}\dots\D{x_k} \,\Delta^2(x_i)\,\prod_{i=1}^{k}(x_0-x_i)^{a}\,\ee{-{N\over T_c}V(x_i)} \cr
&& \qquad \quad \,\,  (n-k)!\,\Zbar_{n-k}(T_c{n-k\over N},{\cal V}_{{r_0},r_j}) \cr
\eea
where
\beq
{\cal V}_{r_j}(x)=   \sum_{j=0}^k r_j \ln{(x_j-x)}
\virg r_0=-{a T_c\over N}\virg r_1=\dots=r_k=-{2 T_c\over N}
\eeq
and $\Zbar_{n-k}(T_c{n-k\over N},{\cal V}_{r_j})$is a one-cut integral:
\bea
\Zbar_{n-k}(T_c{n-k\over N},{\cal V}_{r_j})
&:=& {1\over n-k!}\,\int_{<\td{e}}\,\D{x}_{k+1}\dots \D{x}_{n}\, \Delta^2(x_i)\,\prod_{i=k+1}^{n}\,\ee{-{N\over T_c}{\cal V}_{r_j}(x_i)} \cr
&=& H_{n-k}\, \ee{-{N^2\over T_c^2} \Fbar_{n-k}(T_c{n-k\over N},{\cal V}_{r_j})}
\eea

That gives:
\bea
&& Z_{n}({n\over N}T_c,V_{r_0}) \cr
&\sim& \sum_{k=0}^{n} {H_{n-k}\over k!}
\int_{x_i>\td{e}} \D{x_1}\dots\D{x_k}\,\Delta^2(x_i)\, \prod_{i=1}^k (x_0-x_i)^{a}\,\ee{-{N\over T_c}V(x_i)}
\,\,  \ee{-{N^2\over T_c^2} \Fbar_{n-k}(T_c{n-k\over N},{\cal V}_{r_j})} \cr
\eea
In other words, we integrate out $n-k$ eigenvalues, and consider the integral over $k$ eigenvalues only.
The $k$ remaining eigenvalues are submitted to the potential $V$, as well as their mutual Coulomb repulsion, and
the mean field of the exterior $n-k$ eigenvalues.

Since $\Fbar_{n}(T,{\cal V}_{r_j})$ corresponds to a one-cut distribution,
it can be evaluated with standard semiclassical technique (see section \ref{sectionclassical}), and in particular,
it has a large $n$ expansion:
\beq
\Fbar_{n}(T,{\cal V}_{r_j}) \sim
\Fbar(T,{\cal V}_{r_j}) + {T\over n}\Fbar^{(1/2)}(T,{\cal V}_{r_j}) + {T^2\over n^2}\Fbar^{(1)}(T,{\cal V}_{r_j})
+O({1\over n^3})
\eeq
and each term of the expansion is analytical in ${\cal V}_{r_j}$.

We want to evaluate $\Fbar(T,{\cal V}_{r_j})$
in a regime where $T-T_c$ is "small" (we make that more precise below) and $r_0$ and the
$r_j$'s are of order $O(1/N)$.

we first do a Taylor expansion in $T-T_c$ and the $r_j$'s:
\bea
\Fbar(T,{\cal V}_{r_j}) & \sim &
\Fbar(T_c,V) + \sum_{i=0}^k r_i {\d \Fbar\over \d r_i}
 + {1\over 2} \sum_{i,j}  r_i r_j {\d^2 \Fbar\over \d r_i \d r_j} +o({k^2\over
N^2}) \cr
&&  + \sum_j (T-T_c) r_j {\d^2 \Fbar\over \d T \d r_j} \cr
&& + (T-T_c) {\d \Fbar\over \d T} + {1\over 2} (T-T_c)^2 {\d^2 \Fbar\over \d T^2}
+o({1\over N^2})
\eea
where all derivatives are computed at $T=T_c$ and $r_i=0$ (see section \ref{sectionderivativesT} and \ref{sectionderivativesr}).
That expansion is valid only if $k\ll N$.

Thus we have:
\bea
{1\over H_n}\, Z_{n}({n\over N}T_c,V_{r_0})
& \sim &
\ee{-{N^2\over T_c^2} \Fbar(T_c,V)}
\ee{- \Fbar^{(1)}(T_c,V)}
\ee{a\,{N\over T_c} \Fbar_{r_0}}
\ee{-{a^2\over 2} \Fbar_{r_0,r_0}} \cr
&& \sum_{k=0}^{n} {1\over k!}{H_{n-k}\over H_n}\, \ee{-{N\over T_c} (n-N-k) \Fbar_T} \ee{a (n-N-k) \Fbar_{T,r_0}}
\ee{-{(n-N-k)^2\over 2} \Fbar_{T,T}} \cr
&&  \int_{x_i>\td{e}} \D{x_1}\dots\D{x_k}\,\, \Delta^2(x_i)
\prod_i (x_0-x_i)^{a}\,\ee{-{N\over T_c}V(x_i)}\,
\cr
&& \prod_{j=1}^k \ee{2{N\over T_c} \Fbar_{r_j}}\, \ee{2 (n-N-k) \Fbar_{T,r_j}} \,\ee{-2a \Fbar_{r_0,r_j}}
\prod_{j,l\geq 1} \ee{-2 \Fbar_{r_l,r_j}} \cr
\eea

Notice that:
\beq
{H_{n-k}\over H_n} \sim (2\pi)^{-k} (1-{k\over n})^{-1/12} (1+O(1/n))
\eeq

\subsection{Computation of the derivatives}

Now, use formula given in section \ref{sectionderivativesT} and \ref{sectionderivativesr}
in the one-cut case,
and get (derivatives taken at $T=T_c$ and $r_i=0$, and take into account that
$V_{\rm eff}(b)=V_{\rm eff}(e)$):
\beq
{\d\over \d r_i} \Fbar = -{1\over 2} \left( V_{\rm eff}(x_i) - V(x_i) \right)
\virg
{\d\over \d T} \Fbar = V_{\rm eff}(e)
\eeq
\beq
{\d^2\over \d r_i \d r_j} \Fbar = \ln{x_i-x_j \over \L(x_i)- \L(x_j)}
\virg
{\d^2\over \d r_i^2} \Fbar = -\ln{\L'(x_i)}
\eeq
\beq
{\d^2\over \d T \d r_i} \Fbar = \ln{\L(x_i)}
\virg
{\d^2\over \d T^2} \Fbar = 0
\eeq

Moreover:
\beq
{\d \T(T,V)\over \d T} = {a+b\over 2} =0
\virg
{\d \T(T,V)\over \d r_j} =  {1\over \L(x_j)}
\eeq

The effective potential behaves in the vicinity of $e$ as:
\beq\label{approxVeff}
V_{\rm eff}(x) \mathop{\sim}_{x\to e} V_{\rm eff}(e)+{V_{\rm eff}^{(2\nu)}(e)\over 2\nu!}\,(x-e)^{2\nu}
\eeq
\beq
{V_{\rm eff}^{(2\nu)}(e)\over 2\nu !} = {2\sinh{\phi_e}\,Q(e)\over 2\nu}
\eeq

In the limit where the $x_i$'s are close to $e$, we have:
\beq\label{approxxxLL}
{x_i-x_j \over \L(x_i)- \L(x_j)} \quad \mathop{\sim}_{x_i,x_j\to e} \quad {1\over \L'(x_i)} \sim 2\sinh\phi_e\,\ee{-\phi_e}
\eeq
and
\beq\label{approxL}
\L(x_i)\mathop{\sim}_{x_i\to e} \ee{\phi_e}
\eeq

\subsection{Result}

write $n=N+p$:
\bea
Z_{N+p}({N+p\over N}T_c,V_{r_0})
& \sim &
\ee{-{N^2\over T_c^2} \Fbar(T_c,V)} \ee{- \Fbar^{(1)}(T_c,V)}\cr
&& \left({\ee{\phi_e}\over 2\sinh\phi_e}\right)^{{a^2\over 2}}\,
\ee{-{a N\over 2 T_c} (V_{\rm eff}(x_0)-V(x_0))}  \cr
&& \sum_{k=0}^{N+p} {H_{N+p-k}\over k!}\,
 \ee{-{N\over T_c} (p-k) V_{\rm eff}(e)}\, \ee{a (p-k) \phi_e} \cr
&& \int_{x_i>\td{e}} \D{x_1}\dots\D{x_k}\,\, \Delta^2(x_i)
\prod_{i=1}^k (x_0-x_i)^{a}\,\ee{-{N\over T_c}V(x_i)}\,
\cr
&& \prod_{i=1}^k \ee{-{N\over T_c} (V_{\rm eff}(x_i)-V(x_i))}\,
\ee{2k (p-k) \phi_e} \,
 \left({\ee{\phi_e}\over 2\sinh\phi_e}\right)^{2ak+2k^2} \cr
& \sim &
H_{N}\,\ee{-{N^2\over T_c^2} \Fbar(T_c,V)} \ee{- \Fbar^{(1)}(T_c,V)}\, \ee{-{a N\over 2 T_c} (V_{\rm eff}(x_0)-V(x_0))}\,\ee{-p {N\over T_c} V_{\rm eff}(e)}  \cr
&& \sum_{k=0}^{N+p} {(2\pi)^{p-k}\over k!}\,  \ee{(2k+a) (p-k) \phi_e} \,
 \left({\ee{\phi_e}\over 2\sinh\phi_e}\right)^{2(k+{a\over 2})^2} \cr
&& \int_{x_i>\td{e}} \D{x_1}\dots\D{x_k}\,\, \Delta^2(x_i)
\prod_{i=1}^k (x_0-x_i)^{a}\,\ee{-{N\over T_c}(V_{\rm eff}(x_i)-V_{\rm eff}(e))}\, \cr
\eea
Then we rescale:
\beq\label{scalingxi}
x_i = e + N^{-1\over 2\nu}\, \left({2\sinh\phi_e\,Q(e)\over T_c}\right)^{-1\over 2\nu}\, y_i
\,\, , \,\,
x_0 = e + N^{-1\over 2\nu}\, \left({2\sinh\phi_e\,Q(e)\over T_c}\right)^{-1\over 2\nu}\, y
\eeq
and we get:
\bea\label{asympZgena}
&& Z_{N+p}({N+p\over N}T_c,V_{r_0})\,\ee{-{a N\over 2 T_c} V(x_0)} \cr
& \sim &
H_N\,\ee{-{N^2\over T_c^2} \Fbar(T_c,V)} \ee{- \Fbar^{(1)}(T_c,V)}\, \ee{-(p+a/2){N\over T_c} V_{\rm eff}(e)}\,\ee{-a{y^{2\nu}\over 4\nu}}\cr
&& N^{a^2\over 8\nu}\,(2\sinh\phi_e)^{-a^2\over 2}\,(2\pi)^{p}  \cr
&& \sum_{k=0}^{N+p} N^{-{(k+a/2)^2\over 2\nu}}\,  \ee{(k+a/2) 2p \phi_e} \,\ee{(k+a/2) a\phi_e}\,
 A^{-(k^2+ak)}\,(2\pi)^{-k} \cr
&& {1\over k!}\,\int \D{y_1}\dots\D{y_k}\,\, \Delta^2(y_i)
\prod_{i=1}^k (y-y_i)^{a}\,\ee{-{y_i^{2\nu}\over 2\nu}}\, \cr
&& (1+O(N^{-{1\over 2\nu}}))
\eea
where we have defined:
\beq
A:= (2\sinh\phi_e)^2\,\left({2\sinh\phi_e\,Q(e)\over T_c}\right)^{1\over 2\nu}
\eeq

\eq{asympZgena} is valid only up to $O(N^{-{1\over 2\nu}})$ because \eq{approxVeff}, \eq{approxxxLL} and \eq{approxL} are valid only to that order in the regime of \eq{scalingxi}.

\subsection{The effective matrix model}

Let us define the matrix model in the potential ${y^{2\nu}\over 2\nu}$.

Let $\zeta_{k,\nu}$ be the partition function of the $k\times k$ matrix model in the potential ${y^{2\nu}\over 2\nu}$:
\beq
\zeta_{k,\nu} := {1\over k!}\,\int \D{x_1}\dots\D{x_k}\,\,\Delta^2(x_i) \,\,
\prod_i \ee{-{x_i^{2\nu} \over 2\nu} }\,\,
\eeq
Notice that for $\nu=1$, this is the gaussian matrix model, and we have:
\beq
\ln{\zeta_{k,1}}={k\over 2}\ln{2\pi} + \ln{(\prod_{j=0}^{k-1} j!)}
= \ln{H_k} + {k^2\over 2}\ln{k}-{3\over 4}k^2
\eeq

We define the amplitude:
\beq
A_k := A^{-k^2}\,(2\pi)^{-k}\,\zeta_{k,\nu}
\eeq

We also introduce:
\beq
h_{k,\nu}:={\zeta_{k+1,\nu}\over \zeta_{k,\nu}}= 2\pi\,A^{2k+1}\,{A_{k+1}\over A_k}
\eeq
\beq
\gamma_{k,\nu}:= \sqrt{h_{k,\nu}\over h_{k-1,\nu}} = A\, \sqrt{A_{k+1}A_{k-1}\over A_k^2}
\eeq

and the associated orthogonal polynomials
\beq
P_k(y):=
{\int dy_1\dots dy_k \Delta(y_i)^2 \,\prod_{i=1}^k (y-y_i) \,\ee{-{y_i^{2\nu}\over 2\nu}}
\over \int dy_1\dots dy_k \Delta(y_i)^2 \,\prod_{i=1}^k  \,\ee{-{y_i^{2\nu}\over 2\nu}}}
\eeq
\beq
\ps_{k,\nu}(y):= \sqrt{\zeta_{k,\nu}\over \zeta_{k+1,\nu}}\,P_k(y)\,\ee{-{y^{2\nu}\over 4\nu}}
\eeq
and their Hilbert transforms:
\beq
\hat{P}_{k-1}(y):=
{\int dx_1\dots dx_k \Delta(x_i)^2 \,\prod_{i=1}^k {1\over y-x_i} \,\ee{-{x_i^{2\nu}\over 2\nu}}
\over \int dx_1\dots dx_k \Delta(x_i)^2 \,\prod_{i=1}^k  \,\ee{-{x_i^{2\nu}\over 2\nu}}}
\eeq
\beq
\hatps_{k,\nu}(y) := \sqrt{h_{k,\nu}}\,\hat{P}_{k}(y)\,\ee{y^{2\nu}\over 4\nu}
\eeq

\subsection{Partition function}

Thus we have for $a=0$:
\bea\label{asympZsumgk}
&& Z_{N+p}({N+p\over N}T_c,V) \cr
& \sim&
H_N\,\ee{-{N^2\over T_c^2} \Fbar(T_c,V)} \ee{- \Fbar^{(1)}(T_c,V)}\, \ee{-p{N\over T_c} V_{\rm eff}(e)}\,(2\pi)^{p}
 \sum_{k=0}^{N+p} N^{-{k^2\over 2\nu}}\,  \ee{ 2kp \phi_e} A_k (1+O(N^{-{1\over 2\nu}})) \cr
\eea

\beq\label{asymphsumgk}
h_{N+p}
 \sim
2\pi\, \ee{-{N\over T_c} V_{\rm eff}(e)}\,
{\sum_{k=0}^{N+p+1} N^{-{k^2\over 2\nu}}\,  \ee{ 2kp \phi_e}\,\ee{ 2k \phi_e}\, A_k
\over \sum_{k=0}^{N+p} N^{-{k^2\over 2\nu}}\,  \ee{ 2kp \phi_e}\, A_k} (1+O(N^{-{1\over 2\nu}}))
\eeq

\bea\label{asympgammasumgk}
\gamma_{N+p}^2
& \sim&
{\left(\sum_{k=0}^{N+p+1} N^{-{k^2\over 2\nu}}\,  \ee{ 2kp \phi_e}\,\ee{ 2k \phi_e}\, A_k\right)
\left(\sum_{k=0}^{N+p-1} N^{-{k^2\over 2\nu}}\,  \ee{ 2kp \phi_e}\,\ee{ -2k \phi_e}\, A_k\right)
\over
\left(\sum_{k=0}^{N+p} N^{-{k^2\over 2\nu}}\,  \ee{ 2kp \phi_e}\, A_k\right)^2} \cr
&&  \qquad \quad (1+O(N^{-{1\over 2\nu}}))
\eea

\subsection{Orthogonal polynomial}

According to Heine's formula (cf \eq{deforthoPol}), we have:
$$
\psi_n  = {Z_n(T_c{n\over N},V(x)-{T_c\over N}\ln{(\xi-x))}\over \sqrt{Z_n(T_c{n\over N},V)\, Z_{n+1}(T_c{n+1\over N},V)}} \,\ee{-{N\over 2T}\,V(\xi)}
$$
$$
\phi_{n-1}(\xi) = {Z_{n}(T_c{n\over N},V(x)+{T_c\over N}\ln{(\xi-x))}\over \sqrt{Z_n(T_c{n\over N},V)\,Z_{n-1}(T_c{n-1\over N},V)}} \,\ee{{N\over 2T}\,V(\xi)}
$$

\bea
\psi_{N+p}(x_0) = {Z_{N+p}({N+p\over N}T_c,V_{r_0})\,\ee{-{N\over 2T}V(x_0)}\over \sqrt{Z_{N+p}({N+p\over N}T_c,V)\,Z_{N+p+1}({N+p+1\over N}T_c,V)}} \cr
 r_0=-{T_c\over N}\virg a=1
\eea
thus, in the regime:
\beq
x_0 = e + N^{-{1\over 2\nu}}\,{4\sinh^2\phi_e\over A}\, y
\eeq
 using \eq{asympZgena} with $a=1$ we get:
\bea\label{asymppsisumk}
\psi_{N+p}(x_0)
& \sim &
\, N^{1\over 8\nu}\,\sqrt{A\over 2 \sinh\phi_e}  \cr
&& {\sum_{k=0}^{N+p} N^{-{(k+1/2)^2\over 2\nu}}\,  \ee{(k+1/2) 2p \phi_e} \,\ee{(k+1/2)\phi_e}\, \sqrt{A_k A_{k+1}}\,\,\ps_{k,\nu}(y)
\over \sqrt{(\sum_{k=0}^{N+p+1} N^{-{k^2\over 2\nu}}\,  \ee{ 2kp \phi_e}\,\ee{ 2k \phi_e}\, A_k) (\sum_{k=0}^{N+p} N^{-{k^2\over 2\nu}}\,  \ee{2kp \phi_e} \,A_k)} }
\cr
&& (1+O(N^{-{1\over 2\nu}}))
\eea

\subsection{Hilbert transforms}

According to \eq{deforthoPolCauchy}, we have:
\bea
\phi_{N+p-1}(x_0) = {Z_{N+p}({N+p\over N}T_c,V_{r_0})\,\ee{{N\over 2T}V(x_0)}\over \sqrt{Z_{N+p-1}({N+p-1\over N}T_c,V)\,Z_{N+p}({N+p\over N}T_c,V)}}  \cr
r_0={T_c\over N}\virg a=-1
\eea
thus, in the regime:
\beq
x_0 = e + N^{-{1\over 2\nu}}\,{4\sinh^2\phi_e\over A}\, y
\eeq
using \eq{asympZgena} with $a=-1$ we get:

\bea\label{asympphisumk}
\phi_{N+p-1}(x_0)
& \sim &
 N^{1\over 8\nu}\,\sqrt{A\over 2\sinh\phi_e}  \cr
&& {\sum_{k=0}^{N+p} N^{-{(k-1/2)^2\over 2\nu}}\,  \ee{(k-1/2) 2p \phi_e} \,\ee{-(k-1/2)\phi_e}\,
\sqrt{A_{k-1} A_{k}}\, \hatps_{k-1,\nu}(y)
\over \sqrt{
(\sum_{k=0}^{N+p} N^{-{k^2\over 2\nu}}\,  \ee{ 2kp \phi_e} \, A_k )
(\sum_{k=0}^{N+p-1} N^{-{k^2\over 2\nu}}\,  \ee{ 2kp \phi_e}\,\ee{ -2k \phi_e} \, A_k )
}}
\cr
&& (1+O(N^{-{1\over 2\nu}}))
\eea
Notice that shifting $k\to k+1$ we have:
\bea\label{asympphisumkbis}
\phi_{N+p-1}(x_0)
& \sim &
 N^{1\over 8\nu}\,\sqrt{A\over 2\sinh\phi_e}  \cr
&& {\sum_{k=-1}^{N+p-1} N^{-{(k+1/2)^2\over 2\nu}}\,  \ee{(k+1/2) 2p \phi_e} \,\ee{-(k+1/2)\phi_e}\,
\sqrt{A_{k+1} A_{k}}\, \hatps_{k,\nu}(y)
\over \sqrt{
(\sum_{k=0}^{N+p} N^{-{k^2\over 2\nu}}\,  \ee{ 2kp \phi_e} \, A_k )
(\sum_{k=0}^{N+p-1} N^{-{k^2\over 2\nu}}\,  \ee{ 2kp \phi_e}\,\ee{ -2k \phi_e} \, A_k )
}}
\cr
\eea

\subsection{Computation of $\beta_{N+p}$}

We start from \eq{defbetan}
\bea
{N\over T_c}\, \T_{n}({n\over N}T_c,V)
& \sim & {1\over n!\, Z_n({n\over N}T_c,V)}
\int_{x_i>\td{e}} \D{x_1}\dots\D{x_k}\,\, \ee{-{N\over T_c}V(x_i)} \Delta^2(x_i) \cr
&& (\sum_{j=1}^k x_j+ {N \over T_c}\, \T_{n-k}(T_c{n-k\over N},{\cal V}))
\,\,  \ee{-{N^2\over T_c^2} \Fbar_{n-k}(T_c{n-k\over N},{\cal V})}
\eea
We have:
\bea
\T_{n-k}(T,{\cal V})
& \sim & \T(T_c,V) + (T-T_c) {\d \T\over \d T}(T_c,V)
+ \sum_j r_j {\d \T\over \d r_j}(T_c,V) + O(1/N^2) \cr
& \sim &  - {2T_c\over N} \sum_j \L(x_j)^{-1} + O(1/N^2)
\eea
Therefore:
\bea
{N\over T_c}\, \T_{n}({n\over N}T_c,V)
& \sim & {1\over Z_n({n\over N}T_c,V)}\,
\sum_{k=0}^{n} { H_{n-k}\over k!}
\int_{x_i>\td{e}} \D{x_1}\dots\D{x_k}\,\, \ee{-{N\over T_c}V(x_i)} \Delta^2(x_i) \cr
&& (\sum_{j=1}^k x_j-2\L_j^{-1})
\,\,  \ee{-{N^2\over T_c^2} \Fbar_{n-k}(T_c{n-k\over N},{\cal V})}\cr
& \sim & {2\sinh{\phi_e}\over Z_n({n\over N}T_c,V)}\,
\sum_{k=0}^{n} {H_{n-k}\over k!}
\, k \int_{x_i>\td{e}} \D{x_1}\dots\D{x_k}\,\, \ee{-{N\over T_c}V(x_i)} \Delta^2(x_i) \cr
&&   \ee{-{N^2\over T_c^2} \Fbar_{n-k}(T_c{n-k\over N},{\cal V})} \cr
& \sim & 2\sinh{\phi_e}\, {\sum_k k\, \ee{2pk\phi_e}\,N^{-{k^2\over 2\nu}}\,  A_k \over \sum_k \ee{2pk\phi_e}\,N^{-{k^2\over 2\nu}}\,  A_k}
\eea
and:
\beq\label{asympbetangen}
\beta_{N+p}
\sim
2\sinh{\phi_e}\, \left(
{\sum_k k\, \ee{2(p+1)k\phi_e}\,N^{-{k^2\over 2\nu}}\,  A_k \over \sum_k \ee{2(p+1)k\phi_e}\,N^{-{k^2\over 2\nu}}\,  A_k}
-
{\sum_k k\, \ee{2pk\phi_e}\,N^{-{k^2\over 2\nu}}\,  A_k \over \sum_k \ee{2pk\phi_e}\,N^{-{k^2\over 2\nu}}\,  A_k}
\right)
\eeq


\section{Asymptotic regimes}

Consider $p$ of order $\ln{N}$:
\beq\label{scalingplnN}
p= {u\over 2\nu\phi_e}\ln{N}
\virg u\,\,\hbox{finite}.
\eeq

\subsection{Possible asymptotic regimes for the partition function}

Then \eq{asympZsumgk} becomes:
\beq\label{asympZsumgkpscaling}
\encadremath{
Z_{N+p}({N+p\over N}T_c)
 \sim
H_N\,\ee{-{N^2\over T_c^2} \Fbar(T_c,V)} \ee{- \Fbar^{(1)}(T_c,V)}\, \ee{-p{N\over T_c} V_{\rm eff}(e)}\,(2\pi)^{p}
\,\, \sum_{k=0}^{N+p} N^{{2ku-k^2\over 2\nu}}\, A_k
}\eeq
It is clear that the sum over $k$ is dominated by the values of $k$ for which the exponent of $N$ is  maximal,
i.e. for which $2uk-k^2$ is maximal.
This means that for $u<0$, the sum is dominated by the vicinity of $k=0$,
and for $u\geq 0$, the sum is dominated by the vicinity of $k=u$.
The sum is then well approximated by a few largest terms.

Let us denote $\ubar$ the positive integer closest to $u$:
\beq
\left\{
\begin{array}{ll}
\ubar := [u+1/2] & \qquad {\rm if}\,\, u\geq 0 \cr
\ubar := 0           & \qquad {\rm if}\,\, u\leq 0 \cr
\end{array}
\right.
\eeq
where $[.]$ denotes the integer part.

Define also:
\beq
\begin{array}{ll}
\epsilon_u:={\rm sgn}(u-\ubar)  & \qquad {\rm if}\,\, u> 0 \cr
\epsilon_u:=1                             & \qquad {\rm if}\,\, u\leq 0 \cr
\end{array}
\eeq

We always have:
\beq
\epsilon_u (u-\ubar) \leq {1\over 2}
\eeq

The largest value of $2uk-k^2$ is obtained for $k=\ubar$, and the second largest value is obtained for $k=\ubar+\epsilon_u$.
The difference is:
\beq
(2u \ubar -\ubar^2) - (2u(\ubar+\epsilon_u)-(\ubar+\epsilon_u)^2) = 1-2\epsilon_u(u-\ubar)
\eeq

\begin{figure}[bth]\label{figscieu}
\hrule\hbox{\vrule\kern8pt
\vbox{\kern8pt \vbox{
\begin{center}
{\mbox{\epsfxsize=10.truecm\epsfbox{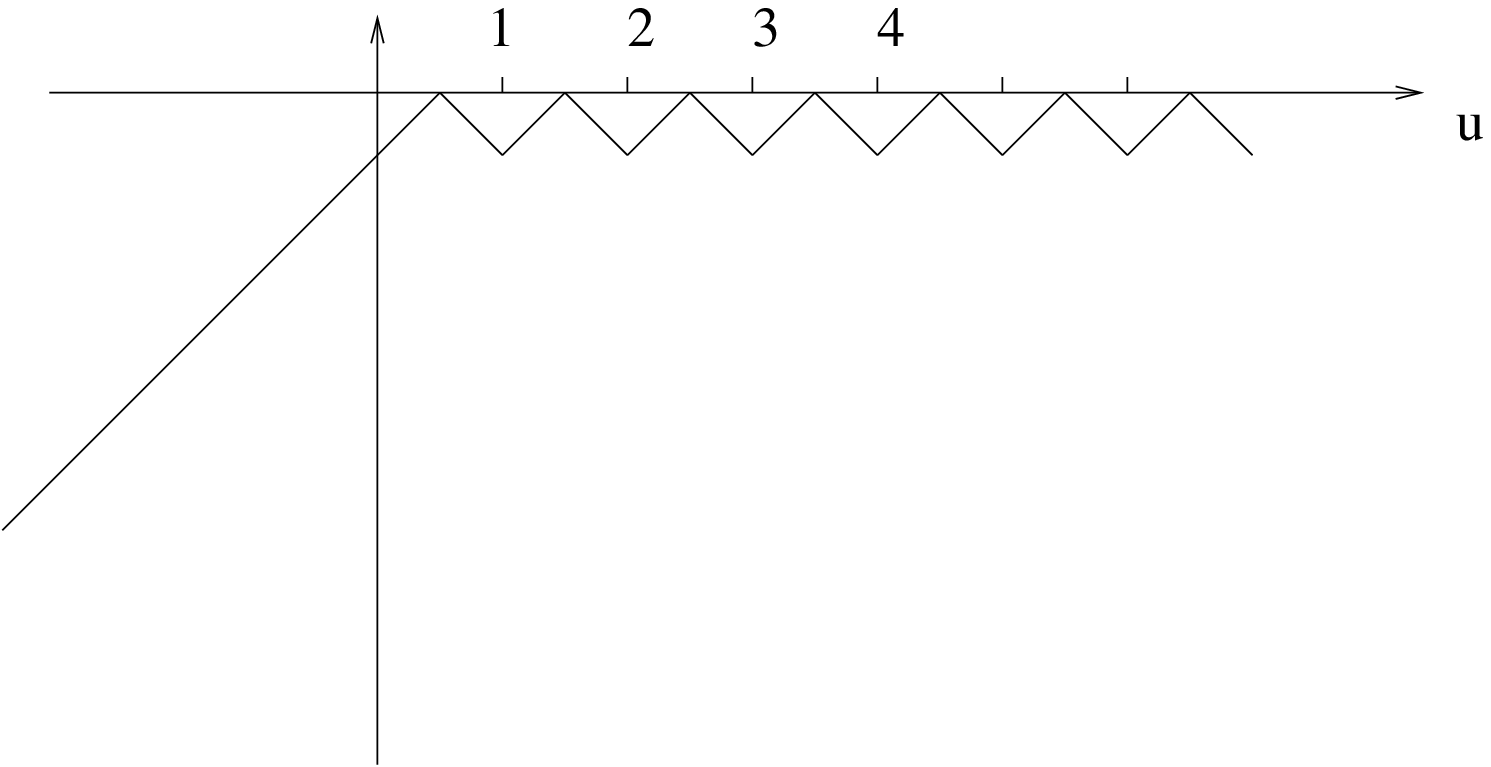}}}
\end{center}
\caption{
Behavior of $\epsilon_u(u-\ubar)-{1\over 2}$.}
}\kern8pt}
\kern8pt\vrule}\hrule
\end{figure}

Remember that our asymptotics for $Z_n$ are valid only up to order $O(N^{-1/2\nu}$, i.e. we want to have:
\beq
\epsilon_u(u-\ubar)-{1\over 2} \geq -{1\over 2}
\eeq
which implies that our asymptotics for $Z_n$ are valid only if $u>0$ and $u\notin {\mathbf N}$.

\subsection{Possible asymptotic regimes for the orthogonal polynomials}

We have similar considerations for the asymptotics of orthogonal polynomials, excpet that the sum over $k$ is now shifted by $1/2$.
This gives different regimes.

\eq{asymppsisumk} becomes in that regime
\bea\label{asymppsisumkas}
\psi_{N+p}(x_0)
& \sim &
\, N^{1\over 8\nu}\,\sqrt{A\over 2 \sinh\phi_e}  \cr
&& {\sum_{k=0}^{N+p} N^{{2(k+1/2) u-(k+1/2)^2\over 2\nu}} \,\ee{(k+1/2)\phi_e}\, \sqrt{A_k A_{k+1}}\,\,\ps_{k,\nu}(y)
\over \sqrt{(\sum_{k=0}^{N+p+1} N^{-{k^2\over 2\nu}}\,  \ee{ 2kp \phi_e}\,\ee{ 2k \phi_e}\, A_k) (\sum_{k=0}^{N+p} N^{-{k^2\over 2\nu}}\,  \ee{2kp \phi_e} \,A_k)} }
\cr
&& (1+O(N^{-{1\over 2\nu}}))
\eea

The sum in the numerator is dominated by the largest values of
\beq
2(k+1/2) u-(k+1/2)^2 ,
\eeq
i.e. by $k=[u]=\ubar+{\epsilon_u-1\over 2}$.

The second largest term is obtained for $k=\ubar-{\epsilon_u+1\over 2}$.
Notice that if $u\geq {1\over 2}$, the two largest terms are always $\ubar$ and $\ubar-1$.
The difference
\beq
(2(\ubar+1/2) u-(\ubar+1/2)^2) - (2(\ubar-1/2) u-(\ubar-1/2)^2)
= 2(u-\ubar)
\eeq

Since our asymptotics are valid only up to order $O(N^{-1\over 2\nu})$, the subleading term should be discarded if $|u-\ubar|\geq {1\over 2}$, i.e. if $u\leq {1\over 2}$ or if $u$ is half-integer.

This implies that our asymptotics for $\psi_n$ are valid only if $u>{1\over 2}$ and $u\notin {\mathbf N}+{1\over 2}$.

\subsection{Asymptotics in the regime $u>0$, and $u$ not integer or half-integer}

From now on, we write:
\beq
n=N+p
\virg
p= {u\over 2\nu\phi_e}\ln{N}.
\eeq

In this section, we assume that $u>0$ and  $u$ not integer or half integer.
The sum over $k$ in \eq{asympZsumgkpscaling} is dominated by the terms $k=\ubar$ and $k=\ubar+\epsilon_u$.

\subsubsection{coefficient $\gamma_n$}

\medskip
Thus we have:
\bea
Z_{N+p}({N+p\over N}T_c)
& \sim &
 H_N  \,\ee{-{N^2\over T_c^2} \Fbar(T_c,V)}
\ee{- \Fbar^{(1)}(T_c,V)} \quad  (2\pi)^p\, N^p\,\ee{-{Np\over T_c} V_{\rm eff}(e)} \,\, \cr
&& \quad  N^{{2u\ubar-\ubar^2\over 2\nu}}\,  \left( A_{\ubar} +  N^{{|u-\ubar|-{1\over 2}\over \nu}}\,  A_{\ubar+\epsilon_u} +O(N^{-{1\over 2\nu}})\right) \cr
\eea

\begin{figure}[bth]\label{figgamman}
\hrule\hbox{\vrule\kern8pt
\vbox{\kern8pt \vbox{
\begin{center}
{\mbox{\epsfxsize=10.truecm\epsfbox{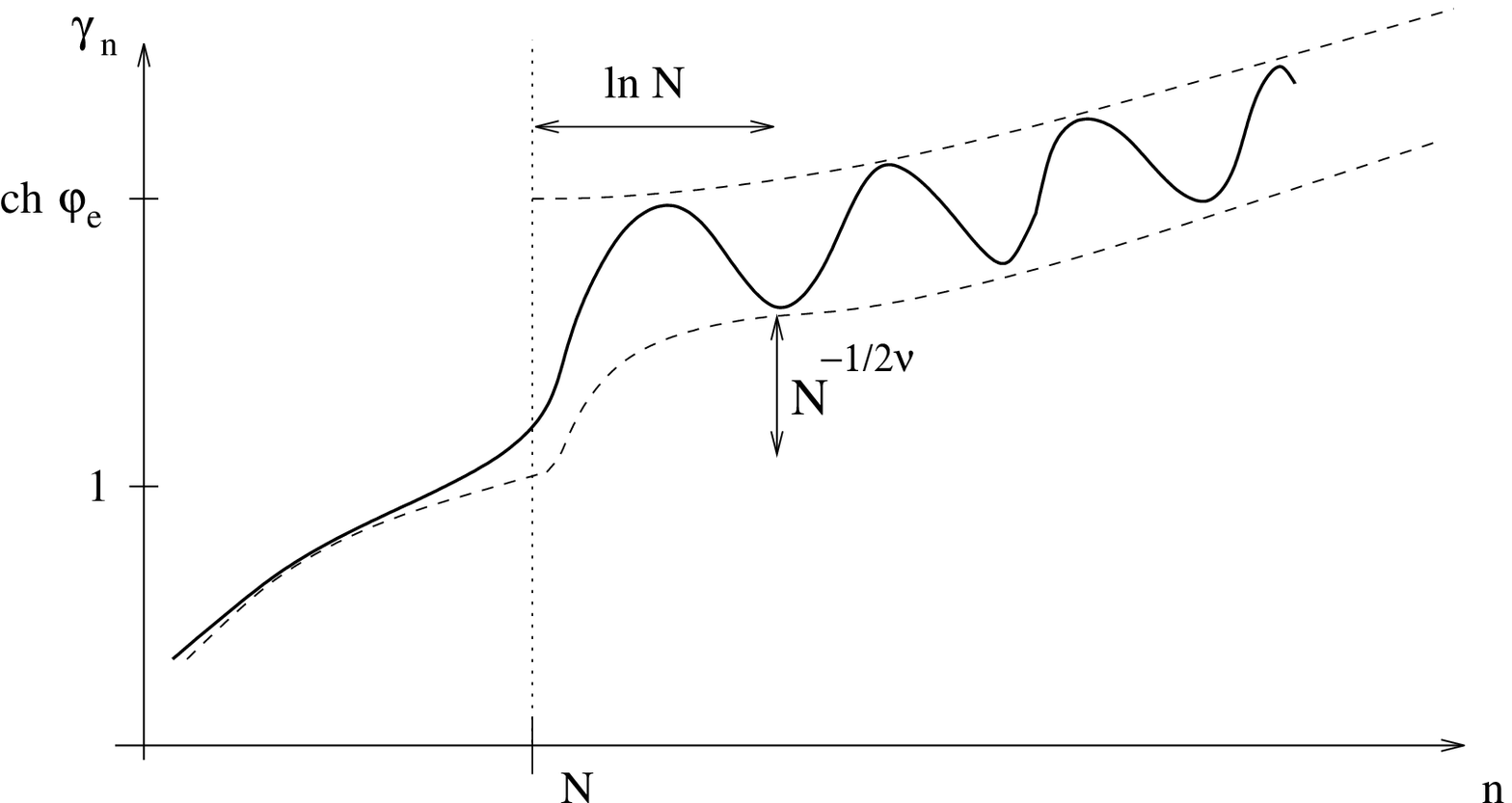}}}
\end{center}
\caption{
Behavior of $\gamma_{n}$.}
}\kern8pt}
\kern8pt\vrule}\hrule
\end{figure}

We also obtain:
\bea
h_{N+p}
& \sim &
2\pi\, \,\ee{-{N\over T_c} V_{\rm eff}(e)} \, \ee{2\ubar\phi_e}\,\left( 1 +  2\epsilon_u \,\ee{\epsilon_u\phi_e}\sinh\phi_e\, N^{{|u-\ubar|-{1\over 2}\over \nu}}\,{A_{\ubar+\epsilon_u}\over A_{\ubar}} +O(N^{-{1\over 2\nu}})\right) \cr
\eea

and:
\bea
\gamma_{N+p}
& \sim &
 1 +  2\,\sinh^2{\phi_e}\, N^{{|u-\ubar|-{1\over 2}\over \nu}}\,{A_{\ubar+\epsilon_u}\over A_{\ubar}} +O(N^{-{1\over 2\nu}}) \cr
\eea
$\gamma_{N+p}$ is nearly periodic, with period ${\ln{N}\over 2\nu\phi_e}$.
The amplitude is minimal of order $N^{-{1\over 2\nu}}$ for $u$ integer, and is maximal of order $1$ for $u$ half-integer, cf fig. \ref{figgamman}.

\subsubsection{coefficient $\beta_n$}

The sum in \eq{asympbetangen} is dominated by $k=\ubar$ and $k=\ubar+\epsilon_u$, i.e.
\beq
\beta_{N+p}
\sim
 4\sinh^2{\phi_e}\, \,\,N^{{2 |u-\ubar|-1\over 2\nu}}\, \ee{\epsilon_u\phi_e}\,  {A_{(\ubar+\epsilon_u)} \over  A_{\ubar}}
\eeq

\subsubsection{Orthogonal polynomials}

The two largest terms in the numerator of \eq{asymppsisumk} correspond to $k=\ubar$ and $k=\ubar-1$ (not necessarily in this order), thus:
\bea
\psi_{N+p}(x_0)
& \sim &
\,  \sqrt{A\over 2 \sinh\phi_e}  \cr
&& {
 N^{u-\ubar\over 2\nu} \,\ee{\phi_e/2}\, \sqrt{ A_{\ubar+1}\over A_{\ubar}}\,\,\ps_{\ubar,\nu}(y)
+ N^{\ubar-u\over 2\nu}\, \ee{-\phi_e/2}\, \sqrt{A_{\ubar-1}\over  A_{\ubar}}\,\,\ps_{\ubar-1,\nu}(y)
\over
1+ \cosh{\phi_e}\,N^{{|u-\ubar|-{1\over 2}\over \nu}}\,\ee{\epsilon_u \phi_e}\, {A_{\ubar+\epsilon_u}\over A_{\ubar}}
}
\cr
&& (1+O(N^{-{1\over 2\nu}}))
\eea

we also have:
\bea
\psi_{N+p-1}(x_0)
& \sim &
\,  \sqrt{A\over 2 \sinh\phi_e}  \cr
&& {
 N^{u-\ubar\over 2\nu} \,\ee{-\phi_e/2}\, \sqrt{ A_{\ubar+1}\over A_{\ubar}}\,\,\ps_{\ubar,\nu}(y)
+ N^{\ubar-u\over 2\nu}\, \ee{\phi_e/2}\, \sqrt{A_{\ubar-1}\over  A_{\ubar}}\,\,\ps_{\ubar-1,\nu}(y)
\over
1+ \cosh{\phi_e}\,N^{{|u-\ubar|-{1\over 2}\over \nu}}\,\ee{-\epsilon_u \phi_e}\, {A_{\ubar+\epsilon_u}\over A_{\ubar}}
}
\cr
&& (1+O(N^{-{1\over 2\nu}}))
\eea

\subsubsection{Hilbert transforms}

Similarly, using \eq{asympphisumkbis}, we find that the Hilbert transform of the orthogonal polynomial $\pi_n$, are asymptoticaly given by:

\bea
\phi_{N+p-1}(x_0)
& \sim &
 \sqrt{A\over 2\sinh\phi_e} \cr
 &&  { N^{{u-\ubar\over 2\nu}}\,  \ee{- \phi_e/2}\,\sqrt{A_{\ubar+1} \over A_{\ubar}}\, \hatps_{\ubar,\nu}(y)
+ N^{-{u-\ubar\over 2\nu}}\, \ee{ \phi_e/2}\, \sqrt{ A_{\ubar-1}\over A_{\ubar}}\, \hatps_{\ubar-1,\nu}(y)
\over  1 + \cosh{\phi_e}\, N^{{2|u-\ubar|-1\over 2\nu}}\, \ee{ -\epsilon_u \phi_e}\,  {A_{\ubar+\epsilon_u}\over A_{\ubar}} }
\cr
\eea
and
\bea
\phi_{N+p}(x_0)
& \sim &
 \sqrt{A\over 2\sinh\phi_e} \cr
 &&  { N^{{u-\ubar\over 2\nu}}\,  \ee{ \phi_e/2}\,\sqrt{A_{\ubar+1} \over A_{\ubar}}\, \hatps_{\ubar,\nu}(y)
+ N^{-{u-\ubar\over 2\nu}}\, \ee{ - \phi_e/2}\, \sqrt{ A_{\ubar-1}\over A_{\ubar}}\, \hatps_{\ubar-1,\nu}(y)
\over  1 + \cosh{\phi_e}\, N^{{2|u-\ubar|-1\over 2\nu}}\, \ee{ \epsilon_u \phi_e}\,  {A_{\ubar+\epsilon_u}\over A_{\ubar}} }
\cr
\eea

\subsubsection{Matrix form}

The matrix:
\beq
\Psi_n(x) = \pmatrix{\psi_{n-1}(x) & \phi_{n-1}(x) \cr \psi_n(x) & \phi_n(x)}
\eeq

is in that regime ($u\geq 0$):
\beq
\encadremath{
\Psi_n(x)
\sim
  \sqrt{A\over 2\sinh\phi_e}\,\,
L^{-1}\,\pmatrix{\ee{{1\over 2}\phi_e} & \ee{-{1\over 2}\phi_e} \cr \ee{-{1\over 2}\phi_e} &  \ee{{1\over 2}\phi_e}}
\, R\,
\pmatrix{\ps_{\ubar-1,\nu}(y) & \hatps_{\ubar-1,\nu}(y) \cr \ps_{\ubar,\nu}(y) & \hatps_{\ubar,\nu}(y)}
}\eeq

where
\beq
R={\rm diag}\,\left(N^{-{u-\ubar\over 2\nu}}\,\sqrt{ A_{\ubar-1}\over A_{\ubar}} , N^{{u-\ubar\over 2\nu}}\,\sqrt{ A_{\ubar+1}\over A_{\ubar}}\right)
\eeq

\beq
L=1+ \cosh{\phi_e}\, N^{{2|u-\ubar|-1\over 2\nu}}\,{A_{\ubar+\epsilon_u}\over A_{\ubar}}\,\pmatrix{ \ee{ -\epsilon_u \phi_e}& 0\cr 0& \ee{ \epsilon_u \phi_e}}
\eeq

\subsubsection{Kernel}

The kernel $K_n(x,x')$ is given by the Christioffel--Darboux formula:
\beq
K_n(x,x') = \gamma_n\,{\psi_n(x)\psi_{n-1}(x')-\psi_n(x')\psi_{n-1}(x)\over x-x'}
\eeq

We find:
\bea
 K_n(x,x')
&\sim& {N^{{1\over 2\nu}}\,  A\over {4\sinh^2\phi_e}}\,\gamma_{\ubar,\nu}\,
 {  \,\ps_{\ubar,\nu}(y)\ps_{\ubar-1,\nu}(y') - \ps_{\ubar-1,\nu}(y)\ps_{\ubar,\nu}(y')
\over
\, (y-y')\,}
\eea
i.e.
\beq
K_n(x,x')\sim K_{\ubar,\nu}(y,y')\,\,{dy\over dx}
\eeq

\subsubsection{Large $u$ limit}

Using Stirling's formula (cf appendix H), we find that for large $u$, all those asymptotics match with the classical limit of section 4.4.

\section{Conclusion}

We have computed the asymptotics of orthogonal polynomials in the birth of a cut critical limit.
This corresponds to the appearance of a new connected component for the support of the eigenvalue density, away from other cuts.

We have found some universal behaviour,  which depends only on the degree $\nu$ of vanishing of the density at the new cut, and $2\cosh\phi_e$ which parametrizes the distance between the new cut and the old cut.
The parametrix near the new cut, is simply the system corresponding to a model matrix model in the potential $x^{2\nu}$, and with $\ubar=[{1\over 2}+2\nu\phi_e {n-N\over \ln{N}}]$ eigenvalues.

This new universal behaviour does not seem to correspond to a conformal field theory (unlike previously known critical behaviours), because the exponent of $N$ is not constant.

It would be interesting to complete the ``physicist's proof'' presented here, with a mathematical one, for instance using Riemann-Hilbert  metods as in \cite{BlIt, dkmvz}.

\subsection*{Aknowledgements}

We would like to thank Pavel Bleher for initiating this work and for many discussions, we also want to thank B. Dubrovin, T. Grava, I. Kostov, H. Saleur, A. Zamolodchikov, for fruitful discussions on those subjects.
This work is partly supported by the Enigma european network MRT-CT-2004-5652, by the ANR project G\'eom\'etrie et int\'egrabilit\'e en physique math\'ematique ANR-05-BLAN-0029-01,
and by the Enrage european network MRTN-CT-2004-005616.

\appendix{Stirling formula and other asymptotics}

Stirling formula:
\beq\label{Stirling}
n! \sim n^n \ee{-n} \sqrt{ 2\pi n} (1+{1\over 12 n} + \dots)
\eeq
from which we deduce:
\beq
\ln{(1\dots (n-1)!)} \sim {1\over 2} n^2 \ln{n} -{3\over 4} n^2 + {n\over
2}\ln{2\pi} - {1\over 12} \ln{n} +O(1)
\eeq
and
\beq\label{appasympHn}
\ln{H_n} \sim n\ln{2\pi} - {\ln{n}\over 12} + \dots
\eeq

Asymptotics of $\zeta_{k,\nu}$. For large $k$ we have:
\beq
\ln{\zeta_{k,\nu}}
\sim {k^2\over 2\nu}\ln{k}-{3\over 4\nu}k^2 + {k\over \nu}\ln{k} + O(k)
\eeq

\appendix{Elliptical functions}

We introduce a few definitions about elliptical functions \cite{elliptical}:
The elliptical sine function $\sn(u,m)$ is defined by the following identity:
\beq
u = \int_0^{\sn(u,m)}{dy\over \sqrt{(1-y^2)(1-my^2)}}
\eeq
The complete integrals are defined by:
\beq
K(m):=\int_0^{1}{dy\over \sqrt{(1-y^2)(1-my^2)}}
\eeq
\beq
K'(m):=K(1-m)=\int_0^{\infty}{dy\over \sqrt{(1+y^2)(1+my^2)}}
\eeq
\beq
E(u,m):=\int_0^{\sn(u,m)} \sqrt{1-my^2\over 1-y^2}\,\D y
\eeq
\beq
E(m):=\int_0^{1} \sqrt{1-my^2\over 1-y^2}\,\D y
\virg
E'(m):=E(1-m)
\eeq

When $m\to 0$ one has:
\bea
K &\sim & {\pi\over 2}\,\left( 1+{m\over 4}+{9m^2\over 64}+\dots+O(m^3) \right) \cr
K' &\sim & \ln{1\over \sqrt{m}}\,\left( 1+{m\over 4}+{9m^2\over 64}+\dots+O(m^3) \right) \cr
E &\sim & {\pi\over 2}\,\left( 1-{m\over 4}-{3m^2\over 64}+\dots+O(m^3) \right) \cr
E' &\sim & 1-{m\over 2}\ln{1\over \sqrt{m}}+O(m^2)  \cr
\eea


\begin{thebibliography}{99}

\bibitem{BI2}
P. Bleher and A. Its,
Double scaling limit in the matrix model: the
Riemann-Hilbert approach,
Preprint, 2002, (arXiv:math-ph/0201003).

\bibitem{BlIt} P. Bleher, A. Its, ``Semiclassical asymptotics of
orthogonal polynomials, Riemann-Hilbert problem,
and universality in the matrix model'' {\em Ann. of Math.}
 (2) {\bf 150}, no. 1, 185--266 (1999).

\bibitem{BlEycrit} P. Bleher, B. Eynard,
''Double scaling limit in random matrix models and a non-linear hierarchy of differential equations'',
 J. Phys. A36 (2003) 3085-3106, xxx, hep-th/0209087.

\bibitem{BDE}
G. Bonnet, F. David, B. Eynard,
Breakdown of universality in multi-cut matrix models,
J.Phys. A: Math. Gen. {\bf 33} (2000) 6739-6768.

\bibitem{BIPZ} E. Brezin, C. Itzykson, G. Parisi, and J. Zuber, Comm. Math. Phys. {\bf 59},   35  (1978).

\bibitem{Cicuta} G.M.Cicuta ``Phase transitions and random matrices'', cond-mat/0012078.

\bibitem{CDM}
 C. Crnkovic, M. Douglas, and G. Moore.
Loop equations and the topological phase of multi-cut matrix models,
Int. J. Mod. Phys. {\bf A7} (1992) 7693-7711.

\bibitem{KazakoVDK} J.M. Daul, V. Kazakov, I.K. Kostov, ``Rational Theories of
2D Gravity from the Two-Matrix Model'', {\em Nucl. Phys.} {\bf B409}, 311-338
(1993), hep-th/9303093.

\bibitem{Matrixsurf} F. David, ``Planar diagrams, two-dimensional lattice gravity and surface models'', {\em Nucl. Phys.} {\bf B 257 [FS14]} 45 (1985).

\bibitem{DeiftBook} P. Deift, {\em Orthogonal Polynomials and Random Matrices: a Riemann-Hilbert
  Approach}, {\em Courant} (New York University Press, ., 1999).

\bibitem{dkmvz} P. Deift, T. Kriecherbauer, K. T. R. McLaughlin,
S. Venakides, Z. Zhou, ``Uniform asymptotics for polynomials
orthogonal with respect to varying exponential weights and
applications to universality questions in random matrix theory'', {\it
Commun. Pure Appl. Math.} {\bf 52}, 1335--1425 (1999).

\bibitem{DDJT}
K. Demeterfi, N. Deo, S. Jain, and C.-I Tan,
Multiband structure and critical behavior of matrix models,
Phys. Rev. D, {\bf 42} (1990) 4105-4122.

\bibitem{ZJDFG} P. Di Francesco, P. Ginsparg, J. Zinn-Justin, ``2D Gravity and
Random Matrices'', {\em Phys. Rep.} {\bf 254}, 1 (1995).

\bibitem{DFMS} Di Ffrancesco P., Mathieu P., S\'en\'echal D., ``Conformal Field Theory''; Corr. 2nd printing, 1998 (Springer-Verlag, Heidelberg, New York, 1997) Graduate Texts in Contemporary Physics.




\bibitem{thDyson} F.J. Dyson, ''Correlations between the eigenvalues of a random
matrix'', Comm. Math. Phys. 19 (1970) 235-250.

\bibitem{EML} N. Ercolani K. Mac Laughlin, ''Asymptotics of the partition function for random matrices via Riemann--Hilber technics and applications to graphical enumeration'',
IMRN 22003, no 14, 755-820.

\bibitem{Farkas} H.M. Farkas, I. Kra, ''Riemann surfaces'' 2nd edition,
Springer Verlag, 1992.

\bibitem{Fay} J. Fay, ''Theta Functions on Riemann Surfaces'', Lectures Notes in Mathematics, Springer Verlag, 1973.

\bibitem{Joh} K. Johansson, On fluctuations of eigenvalues of random hermitian matrices, Duke Math. J. {\bf 91} (1988), 151--204.

\bibitem{Mehta} M.L. Mehta, {\em Random Matrices}, 2nd edition, (Academic Press, New York, 1991).

\bibitem{PeS} V. Periwal and D. Shevitz,
Exactly solvable unitary matrix models: multicritical potentials and  correlations, Nucl. Phys. B {\bf 333} (1990), 731--746.

\bibitem{Sze} G. Szeg\"o. Orthogonal Polynomials, 3rd ed. AMS, Providence, 1967.

\bibitem{TW} C. Tracy and H. Widom,
Introduction to random matrices.
In: Geometric and quantum aspects of integrable systems
(Scheveningen, 1992), Lecture Notes in Phys.
{\bf 424},  Springer-Verlag, Berlin, 1993, 103-130.

\bibitem{elliptical} Special Functions, Wang Z.X., Guo D.R., World Scientific, 1989.












\end{thebibliography}
\end{document}